\documentclass{aa_v7.0}
\usepackage[varg]{txfonts}
\usepackage{longtable}
\usepackage{graphicx}

\newcommand{\kmprs}  {\mbox{\rm km\,s$^{-1}$}}

\newcommand{\feh} {\mbox{\rm [Fe/H]}}

\newcommand{\ch} {\mbox{\rm [C/H]}}
\newcommand{\oh} {\mbox{\rm [O/H]}}
\newcommand{\co} {\mbox{\rm [C/O]}}
\newcommand{\cfe} {\mbox{\rm [C/Fe]}}
\newcommand{\ofe} {\mbox{\rm [O/Fe]}}

\newcommand{\mgfe} {\mbox{\rm [Mg/Fe]}}

\newcommand{\nife} {\mbox{\rm [Ni/Fe]}}

\newcommand{\alphafe} {\mbox{\rm [$\alpha$/Fe]}}
\newcommand{\teff}  {\mbox{$T_{\rm eff}$}}
\newcommand{\logteff} {\mbox{${\rm log}\,T_{\rm eff}$}}
\newcommand{\logg}  {\mbox{{\rm log}\,$g$}}
\newcommand{\turb}  {\mbox{$\xi_{\rm turb}$}}

\newcommand{\CI} {\ion{C}{i}}
\newcommand{\OI} {\ion{O}{i}}
\newcommand{\oI} {\mbox{\rm [{\ion{O}{i}}]}}
\newcommand{\cI} {\mbox{\rm [{\ion{C}{i}}]}}

\newcommand{\TiI} {\ion{Ti}{i}}

\newcommand{\FeI} {\ion{Fe}{i}}
\newcommand{\FeII} {\ion{Fe}{ii}}
\newcommand{\NiI} {\ion{Ni}{i}}

\newcommand{\Mv} {\mbox{$M_V$}}

\newcommand{\Vtotal}   {\mbox{$V_{\rm total}$}}

\newcommand{\VK}{\mbox{($V\!-\!K)$}}

\newcommand{\by}{\mbox{($b\!-\!y)$}}

\def\ltsima{$\; \buildrel < \over \sim \;$}
\def\simlt{\lower.5ex\hbox{\ltsima}}
\def\gtsima{$\; \buildrel > \over \sim \;$}
\def\simgt{\lower.5ex\hbox{\gtsima}}
\begin{document}

\title{Carbon and oxygen abundances in stellar populations
\thanks{Based on observations made with the Nordic Optical Telescope and
on data products from observations made with ESO Telescopes
at the La Silla Paranal Observatory under programs given in 
Table \ref{table:EW1} and in Tables 1 and 2 of Nissen \& Schuster
(2010).}
\fnmsep\thanks{Tables \ref{table:EW1}, \ref{table:EW2}, \ref{table:res1},
and \ref{table:res2} are provided as online material
and are also available at the CDS via anonymous ftp to
{\tt cdsarc.u-strasbg.fr (130.79.128.5a}) or via
{\tt http://cdsarc.u-strasbg.fr/viz-bin/qcat?/A+A/XXX/xxx}.}}

\author{P. E. Nissen \inst{1,2} \and Y. Q. Chen \inst{1} 
\and  L. Carigi \inst{3} \and W. J. Schuster \inst{4} \and G. Zhao \inst{1}}

\institute{Key Laboratory of Optical Astronomy, National Astronomical Observatories,
Chinese Academy of Sciences, Beijing, 100012, China.
\and Stellar Astrophysics Centre, 
Department of Physics and Astronomy, Aarhus University, Ny Munkegade 120, DK--8000
Aarhus C, Denmark.  \email{pen@phys.au.dk}
\and Instituto de Astronom\'{\i}a,  Universidad Nacional Aut\'{o}noma de M\'{e}xico,
AP 70-264, 04510 M\'{e}xico DF, Mexico
\and Observatorio Astron\'{o}mico Nacional, Universidad Nacional Aut\'{o}noma
de M\'{e}xico, Apartado Postal 877, C.P. 22800 Ensenada, B.C., Mexico.}

\date{Received 12 May 2014  / Accepted 17 June 2014}

\abstract
{Carbon and oxygen abundances in stars are important in many fields  of astrophysics
including nucleosynthesis, stellar structure, evolution of galaxies, and formation of
planetary systems. Still, our knowledge of the abundances of
these elements in different stellar populations is uncertain because of
difficulties in observing and analyzing atomic and molecular lines of C  and O.}
{Abundances of C, O, and Fe are determined for F and G main-sequence stars
in the solar neighborhood with metallicities
in the range $-1.6 < \feh < +0.4$ in order to study trends and possible
systematic differences in the C/Fe, O/Fe, and C/O ratios for
thin- and thick-disk stars as well as high- and low-alpha halo stars.
In addition, we investigate if there is any connection between  
C and O abundances in stellar atmospheres and the occurrence of
planets.}
{Carbon abundances are determined from the $\lambda \lambda \, 5052, 5380$ \CI\ lines  
and oxygen abundances from the $\lambda 7774$  \OI\ triplet and
the forbidden \oI\ line at 6300\,\AA . MARCS model atmospheres 
are applied and non-LTE corrections for the \OI\ triplet are included.}
{Systematic differences between high- and low-alpha halo stars and
between thin- and thick-disk stars are seen in the trends of
\cfe\ and \ofe . The two halo populations and thick-disk stars
show the same trend of \co\ versus \oh , whereas the thin-disk stars
are shifted to higher \co\ values. Furthermore, we find some 
evidence of higher C/O and C/Fe
ratios in stars hosting planets than in stars for which no planets
have been detected.}
{The results suggest that C and O in both high- and low-alpha
halo stars and in thick-disk stars are made mainly in massive 
($M > 8 M_{\odot}$) stars,
whereas thin-disk stars have an additional carbon contribution
from low-mass AGB and massive stars of high metallicity
causing a rising trend of the C/O ratio
with increasing metallicity. However, at the highest metallicities
investigated ($\feh \simeq +0.4$), C/O does not exceed 0.8, which seems
to exclude formation of carbon planets if 
proto-planetary disks have the same composition as their parent stars.}


\maketitle

\newpage

\section{Introduction}
\label{sect:introduction} 
Next to hydrogen and helium, carbon and oxygen are the most abundant elements
in the Universe, and their abundances are of high importance in many fields of 
astrophysics, for example stellar age determinations (Bond et al. \cite{bond13}),
chemical evolution of galaxies (Chiappini et al. \cite{chiappini03};
Carigi et al. \cite{carigi05}; Cescutti et al. \cite{cescutti09}), 
and structure of exoplanets  (Bond et al. \cite{bond10}; Madhusudhan \cite{madhusudhan12}).

While it is generally accepted that oxygen is produced by hydrostatic burning
in massive stars and then dispersed to the interstellar medium in SNeII 
explosions (e.g., Kobayashi et al. \cite{kobayashi06}), the origin of carbon
is more uncertain. Both massive ($M > 8 M_{\odot}$) and low-
to intermediate-mass stars probably
contribute, but their relative importance and yields are not well known
due to uncertainties about metallicity-dependent mass loss
(Meynet \& Maeder \cite{meynet02}; van den Hoek \& Groenewegen \cite{hoek97}).

The nucleosynthesis and Galactic evolution of C and O may be studied by
determining abundances in F, G, and K stars with different ages and metallicities.
There is, however, still considerable uncertainty about the abundances
of C and O, because the available atomic and molecular lines
provide diverging results depending on non-LTE corrections and 
atmospheric models applied (Asplund \cite{asplund05}).
In the case of oxygen, all studies show a rising trend of the 
oxygen-to-iron ratio as a function of decreasing iron abundance,
but the derived level of [O/Fe]\,\footnote{For two elements, X and Y,
with number densities $N_{\rm X}$ and $N_{\rm Y}$,
[X/Y] $\equiv {\rm log}(N_{\rm X}/N_{\rm Y})_{\rm star}\,\, - 
\,\,{\rm log}(N_{\rm X}/N_{\rm Y})_{\rm Sun}$.} among metal-poor halo 
star ranges from 0.4\,dex to 0.8\,dex, corresponding to more than a factor of two 
in the O/Fe ratio 
(Nissen et al. \cite{nissen02}; Fulbright \& Johnson \cite{fulbright03};
Cayrel et al. \cite{cayrel04}; Garc\'{\i}a P\'{e}rez \cite{garcia06};
Ram\'{\i}rez et al. \cite{ramirez13}). Carbon, on the other hand, follows iron
more closely than oxygen, but some studies suggest small deviations
from the solar C/Fe ratio in low-metallicity disk and halo stars 
(Gustafsson et al. \cite{gustafsson99}; Shi et al. \cite{shi02};
Reddy et al. \cite{reddy06};
Fabbian et al. \cite{fabbian09}; Takeda \& Takada-Hidai \cite{takeda13}).

Additional information about the origin and Galactic evolution of 
carbon and oxygen may be obtained from differences in \cfe , \ofe , and \co\
between stellar populations.
Precise abundance studies of F and G main-sequence
stars in the solar neighborhood have revealed a clear difference
in \ofe\ between thin- and thick-disk stars in the metallicity range
$-0.7 < \feh < -0.2$ (Bensby et al. \cite{bensby04}, 
Ram\'{\i}rez et al. \cite{ramirez13}). A similar difference in \ofe\
between the two populations of high- and low-alpha\footnote{
alpha refers to the average abundance of the $\alpha$-capture elements,
Mg, Si, Ca, and Ti.} halo stars identified by Nissen \& Schuster
(\cite{nissen10}) has been found by Ram\'{\i}rez et al. (\cite{ramirez12}).  
In the case of carbon, Reddy et al. (\cite{reddy06}) have found evidence
of a systematic difference in \cfe\ between thin- and thick-disk stars,
but this is not confirmed by Bensby \& Feltzing (\cite{bensby06}).
For the high- and low-alpha halo stars, Nissen \& Schuster (\cite{nissen14})
found indications of a systematic difference in \cfe , but this
should be studied further.

More metal-rich stars, i.e., those having $\feh > -0.2$, also seem to 
have a bimodal distribution of various abundance ratios. The precise
abundances of 1111  F and G stars in the solar 
neighborhood based on HARPS spectra
(Adibekyan et al. \cite{adibekyan11}, \cite{adibekyan12})
have revealed the existence of a population of metal-rich, alpha-element-enhanced 
stars having $\alphafe \simeq +0.1$ in contrast to normal thin-disk stars
with $\alphafe \simeq +0.0$. Bensby et al. (\cite{bensby14}) confirm the
existence of these alpha-enhanced stars,  but consider them to belong
to the thick-disk population.
Haywood et al. (\cite{haywood13}) also consider the metal-rich, alpha-enhanced
stars to belong to the thick-disk sequence in the \alphafe\ - \feh\
diagram. They have determined precise ages for the upper 
main-sequence HARPS stars and find the thick-disk stars to have a 
well-defined age-metallicity relation ranging from
13\,Gyr at $\feh \simeq -1$ to 8 Gyr at $\feh \simeq +0.2$. 
In contrast, thin-disk stars with  $\feh > -0.2$ have ages less
than 8\,Gyr and a poorly defined age-metallicity relation.

The analysis of HARPS stars by Adibekyan et al. (\cite{adibekyan12}) 
does not include C and O abundances.
Hence, it remains to be seen if the two disk populations
can be traced in \cfe\ and \ofe\ at metallicities up to 
$\feh \simeq +0.2$. The abundances of C and O are furthermore
of great interest in connection with exoplanets. 
In some recent studies with carbon abundances derived
from high-excitation \CI\ lines and oxygen abundances
from the $\lambda 6300$ \oI\ line (Delgado Mena et al. \cite{delgado10};
Petigura \& Marcy \cite{petigura11}) the carbon-to-oxygen ratio
\footnote{C/O is defined as $N_{\rm C}/N_{\rm O}$,
where $N_{\rm C}$ and $N_{\rm O}$ are the number densities of
carbon and oxygen nuclei, respectively. It should not
be confused with the solar-normalized logarithmic ratio, [C/O]}, C/O,
in metal-rich F and G stars
has been found to range from $\sim \! 0.4$ to $\simgt 1.0$,
i.e., up to a factor of two higher than the solar ratio, 
C/O$_{\odot} \simeq 0.55$ (Asplund et al. \cite{asplund09};
Caffau et al. \cite{caffau08}, \cite{caffau10}).
This has led to suggestions about the existence of exo-planets 
consisting of carbides and graphite instead of Earth-like silicates
(Kuchner \& Seager \cite{kuchner05}; Bond et al. \cite{bond10}).
However, an alternative study by Nissen (\cite{nissen13}) with oxygen abundances
determined from the $\lambda \, 7774$ \OI\ triplet results in a 
tight, slightly increasing relation between C/O and \feh\ corresponding to 
C/O $\simeq 0.8$ at the highest metallicities ($\feh = +0.4$).
This result is supported by a recent study of Teske et al. (\cite{teske14})
of C/O ratios in 16 stars with transiting planets.  Furthermore,
the very low frequency ($< 10^{-3}$) of carbon stars
among K and M dwarf suggests that C/O\,$> 1$ is also very rare among
solar-type stars (Fortney \cite{fortney12}). Still, there is
need for further studies of C/O ratios in stars hosting planets.

There may be other effects in addition to chemical evolution and 
population differences, which cause carbon and oxygen abundances
to vary among F and G stars. In a very precise comparison of abundances
in the Sun and 11
solar twin stars, Mel\'{e}ndez et al. (\cite{melendez.etal09}) found
the Sun to have a $\sim 20$\,\% depletion of refractory elements
like Fe relative to volatile elements like C and O. They 
suggest that this may be related to depletion of refractory elements
when terrestrial planets formed. Their results are supported 
by Ram\'{\i}rez et al. (\cite{ramirez14}), who in a strictly differential
abundance analysis for groups of stars with similar  \teff , \logg , and
\feh\ values find the slope of [X/Fe] versus condensation temperature $T_{\rm C}$
of element X to vary with an amplitude of $\sim \! 10^{-4}$\,dex K$^{-1}$
corresponding to variations of $\sim \! 0.1$\,dex 
in \cfe\ and \ofe .  Based on HARPS spectra, Gonz\'{a}lez Hern\'{a}ndez et al.
(\cite{gonzalez10}, \cite{gonzalez13}) find, however, smaller 
variations of the slope of  [X/Fe] versus  $T_{\rm C}$ and  both
positive and negative slopes for ten stars with detected super-Earth
planets, which casts doubts about the suggestion that a high
volatile-to-refractory element ratio
can be used as a signature of terrestrial planets. Clearly, more
work is needed to explain the  small variations in the 
volatile-to-refractory element ratio and to confirm that such variations
really occur.

In this paper we address some of the problems mentioned above by
determining precise C and O abundances for two samples of F and G stars,
i.e., 66 disk stars with HARPS and FEROS spectra, and 85 
halo and thick-disk stars from Nissen \& Schuster (\cite{nissen10}).
We aim at getting new information on the evolution of \cfe , \ofe ,
and \co\ in stellar populations and to determine C/O ratios for stars
with and without detected planets.

\section{Stellar spectra and equivalent widths}
\label{sect:EWs}

\subsection{The HARPS-FEROS sample of disk stars}
\label{sect:HARPS-FEROS}
Based on Adibekyan et al. (\cite{adibekyan12}) we have selected 
66 main-sequence stars
with 5400\,K $ < \teff < 6400$\,K, which have high signal-to-noise (S/N)
HARPS and FEROS spectra available in the ESO Science Archive.
This sample includes 32 stars with detected
planets from Nissen (\cite{nissen13}), and 34 new stars many of which
have no detected planets. Except for \object{HD\,203608} with  
$\feh = -0.66$, the stars have metallicities in the range $-0.5 < \feh < +0.5$,
and most of them are thin-disk stars, but a few have thick-disk kinematics.

The HARPS spectra cover a wavelength range
from 3800 to 6900\,\AA\ with a resolution of 
$R\simeq \! 115\,000$ (Mayor et al. \cite{mayor03}).
After combination of many individual spectra
of a given star, the S/N exceeds 300 for most of the stars.
These spectra were used to measure the equivalent widths (EWs) of the
$\lambda \lambda 5052, 5380$ \CI\ lines 
and the forbidden [OI] line at 6300\,\AA ,
if not disturbed by telluric O$_2$ lines. In addition, the EWs of 12 \FeII\ lines
listed in Table \ref{table:linedata} were measured from the HARPS spectra.

The FEROS (Kaufer et al. \cite{kaufer99}) spectra, which have a resolution of  
$R\simeq \! 48\,000$ and a typical S/N of 200, were used to measure the EWs of  
the \OI\ triplet lines at 7774\,\AA . 

As described in  Nissen (\cite{nissen13}),
the spectra were first normalized with the IRAF {\tt continuum} task using a low order
cubic spline fitting function. Then the IRAF {\tt splot} task was used to
measure equivalent widths by Gaussian fitting relative 
to local continuum regions selected to be free of lines in the solar spectrum.
Care was taken to use the same continuum windows in all stars.

\begin{table}
\caption[ ]{Line data and derived solar abundances.}
\label{table:linedata}
\setlength{\tabcolsep}{0.10cm}
\begin{tabular}{cccrcccc}
\noalign{\smallskip}
\hline\hline
\noalign{\smallskip}
  ID & Wavelength & $\chi_{\rm exc}$ & log($gf$) & $EW_{\odot}$ &
 $A({\rm X})_{\odot}$\tablefootmark{a} &  $A({\rm X})_{\odot}$  \\
          &  (\AA )    &  (eV)  &   &  (m\AA )  & LTE  & non-LTE \\
\noalign{\smallskip}
\hline
\noalign{\smallskip}
\CI\   & 5052.17 & 7.685 & $-1.301$ & 35.9 & 8.44 & 8.43 \\
\CI\   & 5380.34 & 7.685 & $-1.616$ & 21.5 & 8.44 & 8.43 \\
\oI\   & 6300.31 & 0.000 & $-9.720$ &  3.7\tablefootmark{b} & 8.68 &      \\
\OI\   & 7771.94 & 9.146 &  0.369   & 71.8 & 8.88 & 8.66 \\
\OI\   & 7774.17 & 9.146 &  0.223   & 61.8 & 8.86 & 8.66 \\
\OI\   & 7775.39 & 9.146 &  0.002   & 48.6 & 8.84 & 8.67 \\
\FeII\ & 5414.08 & 3.22  & $-$3.580 & 28.1 & 7.47 &      \\
\FeII\ & 5425.26 & 3.20  & $-$3.220 & 42.0 & 7.43 &      \\
\FeII\ & 5991.38 & 3.15  & $-$3.540 & 31.2 & 7.43 &      \\
\FeII\ & 6084.11 & 3.20  & $-$3.790 & 21.0 & 7.45 &      \\
\FeII\ & 6113.33 & 3.22  & $-$4.140 & 11.6 & 7.47 &      \\
\FeII\ & 6149.25 & 3.89  & $-$2.690 & 36.3 & 7.40 &      \\
\FeII\ & 6238.39 & 3.89  & $-$2.600 & 44.3 & 7.51 &      \\
\FeII\ & 6239.95 & 3.89  & $-$3.410 & 12.7 & 7.41 &      \\
\FeII\ & 6247.56 & 3.89  & $-$2.300 & 52.7 & 7.40 &      \\
\FeII\ & 6369.46 & 2.89  & $-$4.110 & 19.8 & 7.43 &      \\
\FeII\ & 6432.68 & 2.89  & $-$3.570 & 41.3 & 7.45 &      \\
\FeII\ & 6456.39 & 3.90  & $-$2.050 & 63.2 & 7.39 &      \\
\noalign{\smallskip}
\hline
\end{tabular}
\tablefoot{
\tablefoottext{a}{For an element X,  $A({\rm X}) \equiv {\rm log} \, (N_{\rm X}/N_{\rm H}) +12.0$.} \\
\tablefoottext{b}{This value refers to the EW of the [\OI ] line after correction for the \NiI\
contribution to the [\OI ]-\NiI\ $\lambda 6300$ blend. The total EW of the blend is
5.4\,m\AA .}
}

\end{table}

References for the stellar spectra, S/N ratios, and measured EW values
are given in Table \ref{table:EW1}.  In addition, we have used
HARPS and FEROS spectra of reflected sunlight from Ceres and Ganymede to
represent the solar-flux spectrum. The combined HARPS spectrum has
$S/N \simeq 600$ and the FEROS solar-flux spectrum has $S/N \simeq 400$
in the \OI -triplet region. The measured EWs are given in  
Table \ref{table:linedata} and are used in a
differential model-atmosphere analysis of the stars with respect to the Sun.

\onltab{2}{
\clearpage \onecolumn
\begin{longtable}{lccccccccccc}
\caption{\label{table:EW1}ESO observing program numbers, S/N ratios of spectra,
and measured equivalent widths for the HARPS-FEROS sample.} \\
\hline\hline
\noalign{\smallskip}
     &  \multicolumn{2}{c}{HARPS}  &  \multicolumn{2}{c}{FEROS} & \multicolumn{7}{c}
{............................................   $EW$(m\AA )  ...........................................} \\
  ID & Program  &  $S/N$  &  Program & $S/N$ & \CI\ $_{5052}$ &
\CI\ $_{5380}$ & \OI\ $_{7772}$ & \OI\ $_{7774}$ & \OI\ $_{7775}$ & 
[\OI ]+Ni\tablefootmark{a} & [\OI ]\tablefootmark{b}    \\ 
\noalign{\smallskip}
\hline
\noalign{\smallskip}
\noalign{\smallskip}
\endfirsthead
\caption{continued} \\
\hline\hline
\noalign{\smallskip}
     &  \multicolumn{2}{c}{HARPS}  &  \multicolumn{2}{c}{FEROS} & \multicolumn{7}{c}
{............................................   $EW$(m\AA ) ............................................} \\
  ID & Program  &  $S/N$  &  Program & $S/N$ & \CI\ $_{5052}$ &
\CI\ $_{5380}$ & \OI\ $_{7772}$ & \OI\ $_{7774}$ & \OI\ $_{7775}$ &
[\OI ]+Ni\tablefootmark{a} & [\OI ]\tablefootmark{b}    \\
\noalign{\smallskip}
\hline
\noalign{\smallskip}
\noalign{\smallskip}
\endhead
\hline
\endfoot
HD\,1237    &  72.C-0488 &  350 & 60.A-9700 & 200 &  30.9 &  16.8 &  59.1 &  50.2 &  36.4 & 6.0 & 3.5 \\
HD\,4208    &      -     &  200 & 83.A-9011 & 200 &  21.4 &  11.9 &  52.3 &  42.6 &  33.6 &     &     \\
HD\,4307    &      -     &  500 &     -     & 250 &  32.0 &  18.9 &  78.1 &  66.1 &  55.2 & 5.9 & 4.9 \\
HD\,4308    &      -     &  600 & 74.D-0086 & 500 &  25.5 &  14.1 &  68.1 &  58.5 &  42.4 &     &     \\
HD\,14374   &      -     &  350 & 83.A-9011 & 200 &  22.4 &  11.6 &  43.6 &  38.1 &  31.5 & 5.0 & 3.2 \\
HD\,16141   &      -     &  300 & 83.A-9011 & 250 &  45.2 &  27.9 &  86.3 &  75.7 &  58.6 &     &     \\
HD\,20782   &      -     &  900 &     -     & 200 &  33.5 &  19.9 &  72.2 &  59.5 &  49.3 & 5.4 & 4.0 \\
HD\,23079   &      -     &  900 & 84.A-9004 & 200 &  34.1 &  19.7 &  80.8 &  68.5 &  51.8 & 4.2 & 3.2 \\
HD\,28185   &      -     &  500 & 83.A-9011 & 250 &  44.5 &  26.9 &  71.7 &  62.7 &  46.2 & 8.6 & 5.1 \\
HD\,30177   &      -     &  250 & 84.A-9004 & 200 &  48.5 &  31.0 &  72.2 &  62.0 &  53.3 &     &     \\
HD\,30306   &      -     &  500 & 82.C-0446 & 150 &  38.5 &  21.5 &  63.8 &  54.5 &  45.8 & 7.5 & 4.1 \\
HD\,40397   &      -     &  300 & 78.D-0760 & 200 &  29.4 &  16.7 &  61.0 &  51.3 &  36.4 & 6.4 & 4.6 \\
HD\,52265   &      -     &  250 & 80.A-9021 & 350 &  59.1 &  38.9 & 109.2 &  95.6 &  83.1 &     &     \\
HD\,65216   &      -     &  500 & 83.A-9003 & 250 &  22.2 &  12.4 &  54.4 &  50.3 &  35.7 & 4.4 & 3.2 \\
HD\,65907   &  88.C-0011 &  900 & 78.D-0760 & 150 &  35.2 &  19.6 &  89.2 &  75.2 &  60.5 & 5.2 & 4.5 \\
HD\,69830   &  72.C-0488 &  800 & 77.C-0573 & 300 &  23.1 &  12.7 &  42.0 &  37.2 &  28.1 & 5.5 & 3.5 \\
HD\,73256   &      -     & 1100 & 83.A-9003 & 200 &  39.7 &  22.5 &  61.5 &  54.5 &  41.3 & 8.1 & 4.0 \\
HD\,75289   &      -     &  650 & 84.A-9003 & 500 &  57.3 &  36.4 & 108.2 &  95.4 &  78.3 & 5.5 & 3.2 \\
HD\,77110   &  82.C-0212 &  500 & 76.B-0416 & 150 &  18.1 &   9.4 &  60.2 &  52.3 &  39.0 & 5.0 & 4.4 \\
HD\,78538   &  72.C-0488 &  350 & 77.D-0525 & 150 &  27.7 &  15.5 &  67.2 &  60.9 &  41.6 & 4.4 & 3.0 \\
HD\,82943   &      -     &  700 & 84.A-9004 & 300 &  55.0 &  35.4 & 100.2 &  87.2 &  72.8 & 6.9 & 4.1 \\
HD\,89454   &      -     &  900 & 77.D-0525 & 150 &  35.9 &  20.5 &  68.1 &  65.0 &  49.3 & 5.9 & 3.2 \\
HD\,92788   &      -     &  400 & 80.A-9021 & 250 &  49.0 &  30.0 &  81.1 &  70.1 &  58.3 & 8.5 & 4.6 \\
HD\,94151   &      -     &  400 & 77.D-0525 & 200 &  33.0 &  19.3 &  59.4 &  58.0 &  37.8 & 5.7 & 3.3 \\
HD\,96423   &      -     &  550 &     -     & 150 &  38.7 &  22.7 &  74.4 &  60.5 &  50.3 & 6.8 & 4.1 \\
HD\,102365  &  60.A-9036 &  450 & 88.C-0892 & 250 &  23.0 &  13.5 &  54.3 &  45.6 &  33.8 & 4.6 & 3.6 \\
HD\,108147  &  72.C-0488 &  650 & 83.A-9013 & 250 &  55.6 &  35.0 & 114.9 &  98.1 &  80.9 & 4.2 & 2.7 \\
HD\,110668  &      -     &  120 & 77.D-0525 & 150 &  49.0 &  29.7 &  85.5 &  78.6 &  60.6 &     &     \\
HD\,111232  &      -     &  550 &     -     & 250 &  19.6 &  10.5 &  54.8 &  45.5 &  33.0 &     &     \\
HD\,114613  &      -     &  700 & 80.D-2002 & 250 &  50.3 &  30.7 &  83.5 &  76.6 &  63.1 & 9.7 & 6.7 \\
HD\,114729  &      -     &  900 &     -     & 350 &  31.6 &  18.2 &  76.0 &  64.9 &  52.6 & 6.2 & 5.3 \\
HD\,114853  &      -     &  750 & 77.D-0525 & 150 &  22.4 &  12.8 &  58.0 &  47.2 &  36.2 &     &     \\
HD\,115617  &  72.D-0707 &  450 & 86.D-0460 & 150 &  27.3 &  15.7 &  52.7 &  48.7 &  31.4 & 6.1 & 4.1 \\
HD\,117207  &  72.C-0488 &  200 & 83.A-9013 & 200 &  43.8 &  26.5 &  74.3 &  61.1 &  51.5 &     &     \\
HD\,117618  &      -     &  600 &     -     & 300 &  44.8 &  28.1 &  92.1 &  79.4 &  60.2 & 5.0 & 3.4 \\
HD\,125184  &      -     &  850 & 85.C-0557 & 250 &  50.5 &  32.9 &  79.6 &  73.0 &  58.7 & 9.4 & 5.0 \\
HD\,125612  &      -     &  350 & 77.D-0525 & 200 &  45.1 &  27.7 &  88.8 &  76.0 &  62.0 & 7.2 & 4.4 \\
HD\,126525  &      -     &  500 &     -     & 150 &  29.8 &  16.8 &  59.2 &  52.9 &  35.4 & 5.7 & 4.1 \\
HD\,128674  &      -     &  500 & 76.B-0416 & 150 &  17.2 &   8.8 &  45.5 &  39.4 &  24.3 & 3.6 & 2.7 \\
HD\,134664  &      -     &  600 & 77.D-0525 & 150 &  37.6 &  21.8 &  75.3 &  70.5 &  48.8 & 5.5 & 3.7 \\
HD\,134987  &      -     &  550 & 83.A-9013 & 250 &  53.7 &  33.8 &  87.5 &  76.4 &  57.8 & 8.9 & 4.8 \\
HD\,136352  &      -     &  700 & 89.C-0440 & 350 &  26.2 &  15.1 &  65.5 &  56.9 &  43.2 & 5.8 & 5.0 \\
HD\,140901  &      -     &  650 & 78.A-9059 & 200 &  33.2 &  19.2 &  60.1 &  53.5 &  42.8 & 5.7 & 3.1 \\
HD\,145666  &      -     &  550 & 77.D-0525 & 150 &  34.3 &  20.4 &  80.6 &  67.6 &  57.3 & 4.4 & 3.1 \\
HD\,146233  &      -     & 1000 & 78.A-9007 & 300 &  35.8 &  21.2 &  73.1 &  62.2 &  51.3 & 5.5 & 3.6 \\
HD\,157347  &      -     &  450 & 89.C-0440 & 500 &  32.0 &  19.1 &  66.1 &  56.9 &  44.3 & 6.2 & 4.2 \\
HD\,160691  &  73.D-0578 &  600 & 83.A-9013 & 250 &  54.1 &  35.1 &  89.2 &  77.1 &  65.6 & 9.5 & 5.5 \\
HD\,168443  &  72.C-0488 &  600 & 83.A-9003 & 300 &  41.6 &  25.5 &  72.9 &  65.5 &  50.7 & 9.8 & 7.1 \\
HD\,169830  &      -     &  550 & 83.A-9013 & 350 &  70.5 &  46.9 & 139.3 & 123.8 & 102.5 & 6.2 & 4.8 \\
HD\,179949  &      -     &  450 & 85.C-0743 & 450 &  59.7 &  38.2 & 118.5 & 101.9 &  83.4 &     &     \\
HD\,183263  &  75.C-0332 &  500 & 79.A-9013 & 250 &  57.4 &  37.5 &  98.7 &  84.4 &  67.7 &     &     \\
HD\,190248  &  74.D-0380 &  700 & 77.A-9009 & 150 &  52.7 &  31.3 &  74.8 &  68.0 &  56.3 &10.1 & 4.6 \\
HD\,196050  &  72.C-0488 &  950 & 83.A-9011 & 300 &  56.8 &  37.1 &  97.5 &  82.6 &  69.5 & 8.2 & 5.0 \\
HD\,196761  &      -     &  750 & 77.A-9009 & 200 &  17.5 &   8.9 &  36.2 &  30.0 &  21.9 &     &     \\
HD\,202206  &      -     &  700 & 83.A-9011 & 200 &  43.3 &  26.7 &  74.7 &  67.0 &  48.7 & 6.8 & 3.3 \\
HD\,203608  &  77.D-0720 &  600 & 85.A-9027 & 300 &  19.7 &  11.0 &  70.5 &  61.2 &  48.5 & 2.0 & 1.8 \\
HD\,206172  &  72.C-0488 &  200 & 77.C-0192 & 120 &  20.4 &  11.3 &  53.8 &  50.4 &  34.0 &     &     \\
HD\,207129  &      -     &  800 & 60.A-9122 & 300 &  35.7 &  20.7 &  80.7 &  67.4 &  54.0 & 5.0 & 3.6 \\
HD\,210277  &      -     &  900 &     -     & 200 &  41.1 &  25.0 &  65.1 &  57.7 &  44.4 & 8.5 & 4.8 \\
HD\,212301  &  82.C-0312 &  350 & 85.C-0743 & 350 &  58.4 &  37.2 & 112.8 &  97.6 &  81.8 &     &     \\
HD\,213240  &  72.C-0488 &  300 & 83.A-9011 & 200 &  51.9 &  33.5 &  99.1 &  82.5 &  71.1 & 6.9 & 4.7 \\
HD\,216435  &      -     &  700 &     -     & 200 &  61.2 &  40.1 & 108.9 &  96.5 &  77.1 & 7.7 & 4.8 \\
HD\,216777  &      -     &  330 & 77.C-0192 & 120 &  17.2 &   8.9 &  46.5 &  39.3 &  32.9 &     &     \\
HD\,216437  &  80.D-0408 &  700 &     -     & 200 &  36.7 &  21.6 &  52.4 &  47.3 &  37.3 & 8.9 & 5.5 \\
HD\,216770  &  72.C-0488 &  600 &     -     & 120 &  54.9 &  36.9 &  96.3 &  81.8 &  65.7 & 8.5 & 3.5 \\
HD\,222669  &      -     &  650 & 77.D-0525 & 150 &  39.5 &  23.6 &  79.0 &  71.9 &  55.2 & 5.2 & 3.4 \\
\noalign{\smallskip}
\hline

\end{longtable}

\tablefoot{
\tablefoottext{a}{The EW of the [\OI ] + \NiI\ blend at 6300.3\,\AA .} \\
\tablefoottext{b}{The EW of [\OI ] $\lambda 6300.3$ after correcting for the 
contribution of the \NiI\ line.}
}

}

\subsection{The UVES-FIES sample of halo and thick-disk stars}
\label{sect:UVES-FIES}
The second sample, for which we have determined C and O abundances,
consists of high-velocity, F and G main-sequence stars in the solar neighborhood 
from  Nissen \& Schuster (\cite{nissen10}). These stars have metallicities 
in the range $-1.6 < \feh < -0.4$ and most of them belong to the halo population,
i.e. they have space velocities with respect to
the local standard of rest (LSR) larger than 180\,\kmprs , but 16 stars
with thick-disk kinematics are included. Spectra of stars on the southern
sky were acquired from the VLT/UVES archive. They have 
resolutions $R\simeq \! 55\,000$ and S/N ratios
from  250 to 500. Northern stars were 
observed with the FIES spectrograph at the Nordic Optical Telescope
at a resolution of $R\simeq \! 40\,000$ and with $S/N \simeq$ 140 - 200.
Further details are given in Nissen \& Schuster (\cite{nissen10}, Tables 1 and 2).

The UVES and FIES spectra include the \CI\ lines at 5052 and 5380\,\AA , but
the \OI\ triplet at 7774\,\AA\ is not covered. Instead, we have obtained 
equivalent widths for the \OI\ lines from other sources, first
of all Ram\'{\i}rez et al. (\cite{ramirez12}), who measured EWs
for the majority of the  Nissen-Schuster stars based on high resolution spectra
observed with the HET/HRS, Keck/HIRES, and Magellan/MIKE spectrographs.
In addition, we have included EW-values from Nissen \& Schuster
(\cite{nissen97}) for 15 stars with ESO NTT/EMMI spectra, and from 
Nissen et al. (\cite{nissen02}) for six stars with UVES image slicer spectra.
Finally, new EW measurements were carried out for two stars that have FEROS
spectra available. Table \ref{table:EW2} lists the results with mean values 
given, if there is more than one source for a star.

\begin{figure}
\resizebox{\hsize}{!}{\includegraphics{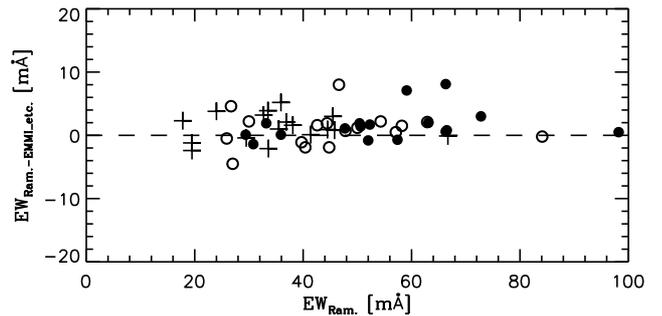}}
\caption{Comparison of \OI\ equivalent widths from Ram\'{\i}rez et al. (\cite{ramirez12})
and values based on EMMI, UVES and FEROS spectra. Filled circles refer to 
\OI\ $\lambda 7771.9$, open circles to \OI\ $\lambda 7774.2$, and crosses to
\OI\ $\lambda 7775.4$.} 
\label{fig:EW-OI}
\end{figure}

As a check of the accuracy of the equivalent width measurements for the \OI\ triplet, 
Fig. \ref{fig:EW-OI} shows a comparison of values from 
Ram\'{\i}rez et al. (\cite{ramirez12}) and the other sources
mentioned (EMMI, UVES, and FEROS). As seen the agreement is quite
satisfactory. There is a mean difference of 1.3\,m\AA\ (Ram\'{\i}rez -- other)
with an rms scatter of 2.4\,m\AA . 

\begin{figure}
\resizebox{\hsize}{!}{\includegraphics{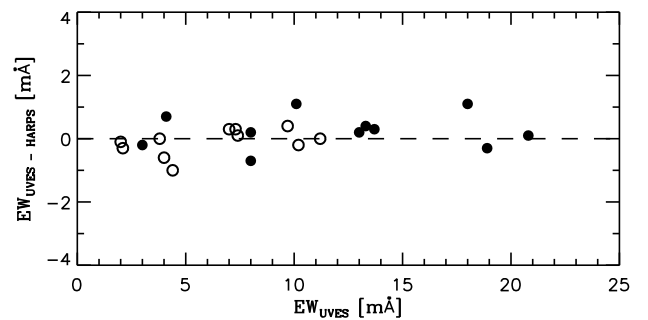}}
\caption{Comparison of \CI\ equivalent widths measured from UVES
and HARPS spectra, respectively. Filled circles refer to
\CI\ $\lambda 5052.2$ and open circles to \CI\ $\lambda 5380.3$.}
\label{fig:EW-CI}
\end{figure}

For 11 stars with UVES spectra, HARPS spectra are also
available. Fig. \ref{fig:EW-CI} shows a comparison of the EWs
of the \CI\ $\lambda \lambda 5052, 5380$ lines measured with the two instruments.
As seen, there in an excellent agreement with a mean difference (UVES -- HARPS)
of 0.1\,m\AA\ and a rms deviation of only 0.5\,m\AA . This good agreement can be
ascribed to the high resolution and S/N of the two sets of spectra.
The FIES spectra have lower resolution and S/N resulting in more uncertain
EW measurements for the \CI\ lines with errors on the order of  2\,m\AA .

The weak [\OI ]-\NiI\ blend could be measured in UVES spectra of 13 of the most
metal-rich halo and thick-disk stars. Five of these 
stars also have HARPS spectra. The mean difference of the two sets
of EWs is 0.0\,m\AA\ and the rms deviation is 0.4\,m\AA . 

\begin{figure}
\resizebox{\hsize}{!}{\includegraphics{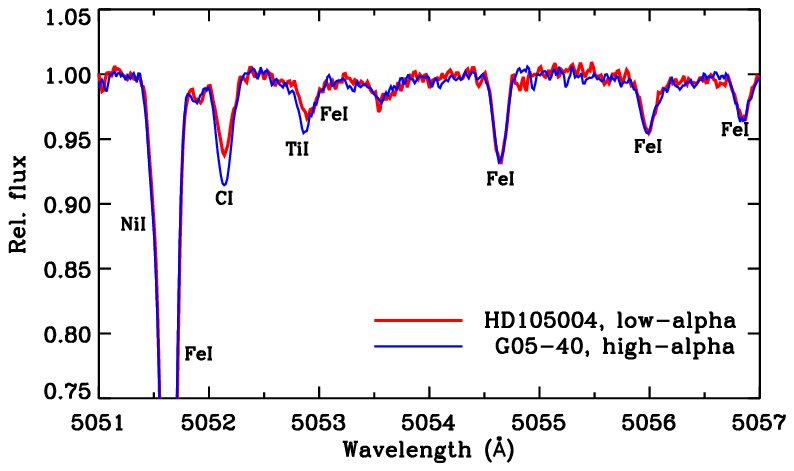}}
\caption{UVES spectra around the $\lambda 5052$ 
\CI\ line for the low-alpha star, HD\,105004 with parameters
(\teff , \logg , \feh , \alphafe ) = (5852\,K, 4.35, $-0.83$, 0.14), 
and the high-alpha star, G\,05-40  with
(\teff , \logg , \feh , \alphafe ) = (5892\,K, 4.20, $-0.83$, 0.31).}
\label{fig:spectra}
\end{figure}

To illustrate the high quality of the spectra applied,
we show in Fig. \ref{fig:spectra} a region around the $\lambda 5052$ 
\CI\ line for a low-alpha star, \object{HD\,105004}, and a high-alpha star,
\object{G\,05-40}, with similar atmospheric parameters and metallicities.
As seen, the \FeI\ lines have the same strengths in the two stars,
but the \CI\ line (and a \TiI\ line) is weaker in the low-alpha star. 

\onltab{3}{
\onecolumn
\begin{longtable}{lcccccccccc}
\caption{\label{table:EW2}Equivalent widths for the UVES-FIES sample of halo and
thick-disk stars.} \\
\hline\hline
\noalign{\smallskip}
    & \multicolumn{2}{c}{..... $EW$(m\AA ) .....} &
    & \multicolumn{3}{c}{............ $EW$(m\AA ) ............} &   
    & \multicolumn{2}{c}{..... $EW$(m\AA ) .....} &   \\  
  ID & \CI\ $_{5052}$ & \CI\ $_{5380}$ & ref.\tablefootmark{a} & \OI\ $_{7772}$ & \OI\ $_{7774}$ & 
  \OI\ $_{7775}$  & ref.\tablefootmark{b} & [\OI ]+Ni\tablefootmark{c} & [\OI ]\tablefootmark{d} &
  ref.\tablefootmark{e} \\
\noalign{\smallskip}
\hline
\noalign{\smallskip}
\noalign{\smallskip}
\endfirsthead
\caption{continued} \\
\hline\hline
\noalign{\smallskip}
    & \multicolumn{2}{c}{.... $EW$(m\AA ) ....} &
    & \multicolumn{3}{c}{............ $EW$(m\AA ) ............} &   
    & \multicolumn{2}{c}{..... $EW$(m\AA ) .....} &   \\  
  ID & \CI\ $_{5052}$ & \CI\ $_{5380}$ & ref.\tablefootmark{a} & \OI\ $_{7772}$ & \OI\ $_{7774}$ & 
  \OI\ $_{7775}$  & ref.\tablefootmark{b} & [\OI ]+Ni\tablefootmark{c} & [\OI ]\tablefootmark{d} &
  ref.\tablefootmark{e} \\
\noalign{\smallskip}
\hline
\noalign{\smallskip}
\noalign{\smallskip}
\endhead
\hline
\endfoot
  BD$-$21 3420 &     7.4 &     4.1 &   U   &    49.4 &    43.4 &    34.6 &   R,E  & 2.2  & 2.1  & U   \\
  CD$-$33 3337 &     7.2 &     3.7 &   U   &    51.5 &    42.6 &    31.6 &   R,E  &      &      &     \\
  CD$-$43 6810 &    33.7 &    18.6 &   U   &    94.7 &    81.6 &    65.0 &   R    & 5.1  & 4.6  & U   \\
  CD$-$45 3283 &     3.2 &     2.0 &   U   &    32.2 &    26.1 &    20.1 &   R,E  &      &      &     \\
  CD$-$51 4628 &     3.3 &     2.4 &   U   &    41.5 &    33.2 &    23.8 &   R    &      &      &     \\
  CD$-$57 1633 &     7.9 &     3.8 &   U,H &    47.1 &    40.2 &    29.7 &   R,E  &      &      &     \\
  CD$-$61 0282 &     3.0 &     1.5 &   U   &    35.8 &    28.9 &    22.1 &   R,E  &      &      &     \\
      G05-19 &     3.7 &     2.1 &   U   &    40.1 &    33.1 &    23.1 &   R    &      &      &     \\
      G05-36 &     6.8 &         &   F   &    52.0 &    45.0 &    39.0 &   R    &      &      &     \\
      G05-40 &    15.0 &     8.0 &   U   &    66.1 &    57.5 &    44.0 &   R,E  &      &      &     \\
      G15-23 &     3.0 &         &   F   &    21.9 &    18.8 &    13.5 &   R    &      &      &     \\
      G18-28 &     7.3 &         &   U   &    34.9 &         &    22.2 &   R    &      &      &     \\
      G18-39 &     4.8 &     2.1 &   U   &    41.8 &    36.9 &    25.7 &   U    &      &      &     \\
      G20-15 &     3.0 &     1.5 &   U   &    30.4 &    27.7 &    16.9 &   R    &      &      &     \\
      G21-22 &     6.2 &     1.9 &   F   &         &         &         &        &      &      &     \\
      G24-13 &    11.8 &     6.6 &   F   &    62.8 &    56.5 &    42.0 &   R    &      &      &     \\
      G31-55 &     5.2 &     3.2 &   F   &    40.4 &    32.8 &    22.3 &   R    &      &      &     \\
      G46-31 &    12.3 &     6.3 &   U   &    58.0 &    50.3 &    36.9 &   E    & 2.5  & 2.3  & U   \\
      G49-19 &    23.0 &    11.5 &   F   &         &         &         &        &      &      &     \\
      G53-41 &     2.6 &     1.6 &   F   &         &    21.8 &    14.1 &   R    &      &      &     \\
      G56-30 &     6.4 &     3.3 &   F   &         &         &         &        &      &      &     \\
      G56-36 &    10.4 &     4.8 &   F   &    58.6 &    50.0 &    38.5 &   R    &      &      &     \\
      G57-07 &    20.8 &    11.6 &   F   &         &         &         &        &      &      &     \\
      G63-26 &     4.5 &     1.9 &   U   &         &         &         &        &      &      &     \\
      G66-22 &     2.3 &     1.3 &   U   &    20.0 &    18.6 &    10.1 &   R    &      &      &     \\
      G74-32 &    15.3 &     7.7 &   F   &         &         &         &        &      &      &     \\
      G75-31 &     7.2 &         &   F   &    55.7 &    44.4 &    34.3 &   U    &      &      &     \\
      G81-02 &    18.3 &     9.2 &   F   &         &         &         &        &      &      &     \\
      G82-05 &     5.4 &     2.3 &   U   &    24.0 &    20.0 &    13.0 &   R    &      &      &     \\
      G85-13 &    13.7 &     7.8 &   F   &    59.0 &    49.1 &    39.5 &   R    &      &      &     \\
      G87-13 &     6.4 &     1.9 &   F   &         &         &         &        &      &      &     \\
      G96-20 &    20.9 &    11.5 &   F   &         &         &         &        &      &      &     \\
      G98-53 &     9.9 &     5.3 &   F   &         &         &         &        &      &      &     \\
      G99-21 &    12.7 &     6.9 &   F   &    50.7 &    39.1 &    31.8 &   R    &      &      &     \\
     G112-43 &     6.7 &     3.3 &   U   &    47.2 &    40.0 &    28.0 &   R    &      &      &     \\
     G112-44 &     3.5 &     2.4 &   U   &         &         &         &        &      &      &     \\
     G114-42 &     3.2 &     2.0 &   U   &         &         &         &        &      &      &     \\
     G119-64 &     3.5 &         &   F   &    37.9 &    34.3 &    22.8 &   R    &      &      &     \\
     G121-12 &     6.4 &     3.2 &   U   &         &         &         &        &      &      &     \\
     G127-26 &    24.1 &    13.3 &   F   &         &         &         &        &      &      &     \\
     G150-40 &     6.8 &     4.7 &   F   &    52.8 &    48.0 &    38.0 &   R    &      &      &     \\
     G159-50 &     8.6 &     4.3 &   U   &    41.9 &         &    24.7 &   R    &      &      &     \\
     G161-73 &     7.0 &         &   F   &         &         &         &        &      &      &     \\
     G170-56 &    10.4 &     5.1 &   F   &    58.9 &    50.1 &    36.9 &   R    &      &      &     \\
     G176-53 &         &         &       &    21.9 &         &    10.4 &   R    &      &      &     \\
     G180-24 &     5.5 &     3.3 &   F   &    44.1 &    38.0 &    29.0 &   R    &      &      &     \\
     G187-18 &    13.8 &     7.8 &   F   &         &         &         &        &      &      &     \\
     G188-22 &     6.0 &     2.8 &   U   &    53.0 &    43.3 &    33.1 &   R    &      &      &     \\
      HD3567 &     6.1 &     3.0 &   U   &    49.9 &    41.8 &    31.1 &   R,U  &      &      &     \\
     HD17820 &    19.3 &    10.1 &   U   &    71.3 &    61.9 &    45.4 &   R,E  & 4.0  & 3.6  & U   \\
     HD22879 &    13.0 &     7.0 &   U,H &    62.1 &    53.2 &    41.4 &   R,F  &      &      &     \\
     HD25704 &    13.6 &     7.4 &   U,H &    62.3 &    49.4 &    37.4 &   R,E  & 2.6  & 2.4  & U,H \\
     HD51754 &    20.8 &    11.2 &   U,H &    70.4 &    62.7 &    44.3 &   R    & 3.9  & 3.4  & U,H \\
     HD59392 &     3.4 &     1.2 &   U   &    37.0 &    28.5 &    19.1 &   R    &      &      &     \\
     HD76932 &    13.4 &     7.3 &   U   &    66.0 &    56.9 &    44.1 &   R,U  & 3.5  & 3.3  & U   \\
     HD97320 &     9.6 &     4.9 &   U,H &    55.6 &    45.7 &    35.9 &   R,F  &      &      &     \\
    HD103723 &    11.0 &     5.9 &   U   &    57.8 &    49.7 &    37.9 &   U    &      &      &     \\
    HD105004 &    10.3 &     5.5 &   U   &    45.9 &    37.2 &    28.8 &   U    &      &      &     \\
    HD106516 &    24.7 &    12.9 &   U   &    98.0 &    84.2 &    66.7 &   R,E  &      &      &     \\
    HD111980 &     8.9 &     5.1 &   U   &    62.0 &    51.6 &    38.4 &   R    & 2.9  & 2.7  & U   \\
    HD113679 &    17.5 &     9.5 &   U,H &    69.3 &    60.0 &    44.8 &   R,E  & 4.9  & 4.4  & U,H \\
    HD114762A&    18.7 &    10.1 &   U   &    70.4 &    60.7 &    46.6 &   R    &      &      &     \\
    HD120559 &     6.6 &     3.5 &   U   &    31.5 &    29.2 &    20.7 &   R,E  & 3.7  & 3.4  & U   \\
    HD121004 &    13.1 &     7.2 &   U,H &    57.7 &    47.5 &    35.0 &   R,E  & 3.8  & 3.4  & U,H \\
    HD126681 &     3.1 &     2.1 &   U,H &    29.4 &    24.4 &    16.7 &   R,E  &      &      &     \\
    HD132475 &     3.8 &     1.7 &   U   &    38.9 &    29.7 &    21.2 &   R    &      &      &     \\
    HD148816 &    19.1 &    10.3 &   U,H &    73.3 &    62.0 &    46.3 &   R    &      &      &     \\
    HD159482 &    14.4 &     7.9 &   F   &    64.7 &    54.9 &    43.6 &   R    &      &      &     \\
    HD160693 &    20.7 &    11.1 &   F   &    72.0 &    60.2 &    45.7 &   R    &      &      &     \\
    HD163810 &     2.1 &         &   U   &    20.0 &         &    12.2 &   R    &      &      &     \\
    HD175179 &    17.6 &     9.6 &   U   &    70.3 &    64.3 &    42.3 &   R    & 4.3  & 3.9  & U   \\
    HD179626 &     9.5 &     5.4 &   U   &    59.0 &    50.2 &    39.6 &   R    &      &      &     \\
    HD189558 &     6.6 &     3.2 &   U   &    51.7 &    41.5 &    30.1 &   R    &      &      &     \\
    HD193901 &     3.8 &     2.2 &   U,H &    32.4 &    26.5 &    18.9 &   R    &      &      &     \\
    HD194598 &     5.7 &     2.6 &   U   &    45.6 &    38.2 &    27.6 &   R    &      &      &     \\
    HD199289 &     8.4 &     4.3 &   U,H &    52.0 &    44.3 &    33.1 &   R    & 2.4  & 2.2  & U,H \\
    HD205650 &     5.7 &     3.1 &   U   &    42.2 &    32.4 &    24.4 &   R    &      &      &     \\
    HD219617 &     2.1 &         &   U   &    30.3 &    26.9 &    17.9 &   R    &      &      &     \\
    HD222766 &     9.9 &     6.6 &   U   &    42.9 &    37.2 &    25.0 &   R    &      &      &     \\
    HD230409 &     4.7 &         &   F   &    27.0 &    23.0 &    18.0 &   R    &      &      &     \\
    HD233511 &         &         &       &    33.2 &    30.1 &    18.3 &   R    &      &      &     \\
    HD237822 &    20.9 &    12.8 &   F   &         &         &         &        &      &      &     \\
    HD241253 &     7.5 &     3.8 &   U   &    52.4 &    41.3 &    33.3 &   R,E  &      &      &     \\
    HD250792A&         &         &       &    29.7 &    22.9 &         &   R    &      &      &     \\
    HD284248 &     2.8 &     1.7 &   U   &    30.9 &    26.3 &    18.1 &   R    &      &      &     \\
\noalign{\smallskip}
\hline

\end{longtable}

\tablefoot{
\tablefoottext{a}{Reference for EWs of \CI\ lines: U, UVES; F, FIES; H, HARPS.} \\
\tablefoottext{b}{Reference for EWs of \OI\ lines: R, Ramirez et al. (\cite{ramirez12});
E, EMMI; U, UVES; F, FEROS.} \\
\tablefoottext{c}{The EW of the [\OI ] + \NiI\ blend at 6300.3\,\AA .} \\
\tablefoottext{d}{The EW of [\OI ] $\lambda 6300.3$ after correcting for the
contribution of the \NiI\ line.} \\
\tablefoottext{e}{Reference for EWs of the \oI\ line: U, UVES; H, HARPS.}}

}

\section{Model-atmosphere analysis and non-LTE corrections}
\label{sect:analysis}

Plane parallel (1D) model atmospheres interpolated 
to the \teff , \logg , \feh , and \alphafe\ values of
the stars were obtained from the 
MARCS grid (Gustafsson et al. \cite{gustafsson08}) and
the Uppsala program EQWIDTH was used to calculate
equivalent widths as a function of element abundance 
assuming LTE. By interpolating to the observed EW, we then
obtain the LTE abundance corresponding to a given line. 

Line data used in the analysis are given in Table \ref{table:linedata}.
The sources of $gf$-values are 
Hibbert et al. (\cite{hibbert93}) for \CI\ lines,   
Hibbert et al. (\cite{hibbert91}) for \OI\ lines, and 
Mel\'{e}ndez \& Barbuy (\cite{melendez09}) for \FeII\ lines.
As the analysis is made differentially to the Sun line by line,
possible errors in
these $gf$-values cancel out. Doppler broadening due to microturbulence
is specified by a depth-independent parameter, $\xi_{\rm turb}$.
Collisional broadening caused by neutral hydrogen
and helium atoms is based on the Uns{\"o}ld (\cite{unsold55}) approximation with an enhancement
factor of two for the \CI\ lines, whereas quantum mechanical calculations of 
Barklem et al. (\cite{barklem00}) and Barklem \& Aspelund-Johansson (\cite{barklem05})
is used for the \OI\ and \FeII\ lines.

Non-LTE corrections calculated by Fabbian et al. (\cite{fabbian09}) were applied to the
oxygen abundances derived from the \OI\ triplet. Fabbian et al.
use a model atom with 54 energy levels and adopt
electron collision cross sections from Barklem (\cite{barklem07}).
Inelastic collisions with hydrogen atoms were described by the
classical Drawin formula (Drawin \cite{drawin68}) scaled by an empirical
factor $S_{\rm H}$. Calculations were performed for both $S_{\rm H}$ = 0 and 1,
which allow us to interpolate the non-LTE corrections to 
$S_{\rm H}$ = 0.85, i.e., the value preferred by  
Pereira et al. (\cite{pereira09}) based on a study of the
solar center-to-limb variation of the \OI\ triplet lines.

As seen from Table  \ref{table:linedata}, the solar non-LTE corrections range 
from $-0.22$ to $-0.17$\,dex for the three \OI\ lines. 
According to Fabbian et al.  (\cite{fabbian09}), the corrections
depend strongly on  \teff ; stars cooler than the Sun have less negative
non-LTE corrections, and  warmer stars have more negative corrections. 
This leads to very significant differential
non-LTE corrections of \oh\  for our sample of stars ranging from about $-0.2$\,dex at 
$\teff \simeq 6300$\,K to +0.1\,dex at $\teff \simeq 5400$\,K.

The non-LTE corrections for carbon abundances derived from the
\CI\ $\lambda \lambda \, 5052 ,5380$ lines
are much smaller than those for the \OI\ triplet. In this paper
we adopt the calculations of Takeda \& Honda (\cite{takeda05}),
who assume $S_{\rm H} = 1$.
The corrections are $-0.01$\,dex for the Sun and changes only slightly 
with \teff\ and \logg . 

Furthermore, we note that the Fe abundance determined from \FeII\ lines
is unaffected by departures from LTE (e.g.,
Mashonkina et al. \cite{mashonkina11}; Lind et al. \cite{lind12}).
This is also the case for oxygen abundances derived from the forbidden \oI\ line at 
6300\,\AA\  (e.g. Kiselman \cite{kiselman93}). This line is, however, affected
by a \NiI\ blend as further discussed in Sect. \ref{sect:COabundances}.

\section{Stellar parameters}
\label{sect:parameters}

\subsection{The HARPS-FEROS sample}
For these stars, Adibekyan et al. (\cite{adibekyan12}) have given 
effective temperatures, \teff , and surface gravities, \logg . 
The values were originally derived from the HARPS spectra by 
Sousa et al. (\cite{sousa08}, \cite{sousa11a}, \cite{sousa11b})
by requesting that \feh\ should not have any systematic dependence
on excitation potential of the lines and that the same iron
abundance is obtained from \FeI\ and \FeII\ lines.

As an alternative, we have determined photometric parameters
for the HARPS-FEROS sample of disk stars.
\teff\ is  derived 
from the \by\ and \VK\ color indices using the calibrations
of Casagrande et al. ({\cite{casagrande10}), which are based on \teff\
values determined with the infrared flux method (IRFM). $V$ magnitudes and
\by\ were taken from Olsen (\cite{olsen83}) and $K$ magnitudes  from the 2MASS catalogue
(Skrutskie et al. \cite{skrutskie06}). For nine stars, the 2MASS
$K$ value is uncertain due to saturation. In these cases, \teff\ has been
determined from \by\ alone. Otherwise we have adopted the mean 
\teff\ derived from \by\ and \VK .

The surface gravity was determined from the relation
\begin{eqnarray} 
\log \frac{g}{g_{\odot}}  =  \log \frac{M}{M_{\odot}} +
4 \log \frac{\teff}{T_{\rm eff,\odot}} + 0.4 (M_{\rm bol} - M_{{\rm bol},\odot}),
\end{eqnarray} 
where $M$ is the mass of the star and $M_{\rm bol}$ the absolute
bolometric magnitude. Hipparcos parallaxes (van Leeuwen \cite{leeuwen07}) are 
used to derive \Mv . Bolometric corrections were adopted from
Casagrande et al. ({\cite{casagrande10}), and
stellar masses were obtained by interpolating in the luminosity - \logteff\
diagram between the Yonsei -Yale evolutionary tracks of Yi et al. (\cite{yi03});
see Nissen \& Schuster (\cite{nissen12}) for details.

Metallicities were determined from the twelve \FeII\ lines listed
in Table \ref{table:linedata}. 
These lines have equivalent widths spanning a range that
allows us to  determine the microturbulence parameter, \turb  ,from the requirement
that the derived \feh\ should not depend on $EW$.
The analysis is made differentially with respect to the Sun
adopting a solar microturbulence of 1.0\,\kmprs .

This procedure of determining stellar parameters
has to be iterated until consistency, because the \teff\ calibrations,
bolometric corrections and mass determinations depend on \feh .

All stars have distances less than 60\,pc
according to their Hipparcos parallaxes. We may therefore assume that
the observed \by\ and \VK\ indices and hence \teff\
are not affected by interstellar reddening.

The rms dispersion of the difference of \teff\ determined
from \by\ and \VK , respectively, is 62\,K. From this
we estimate that the mean photometric temperature, 
\teff (phot) = 1/2 $\times$ (\teff \by\ + \teff \VK)
is determined with an internal one-sigma error 
of $\sim \! 30$\,K. At a given metallicity,
this high precision is confirmed by comparing the photometric
\teff\ values  with the spectroscopic temperatures of 
Adibekyan et al. (\cite{adibekyan12}),
who also quote an error of 30\,K; the rms scatter of 
$\Delta \teff = \teff ({\rm phot}) - \teff ({\rm spec})$
in metallicity intervals of 0.2\,dex is about 40\,K. As seen from
Fig. \ref{fig:Teff.compare}, there is, however, a systematic trend of $\Delta \teff$ 
as a function of metallicity. The difference between the two sets of temperatures 
decreases from about +50\,K at $\feh = -0.4$  to  about $-60$\,K at $\feh = +0.4$.

\begin{figure}
\resizebox{\hsize}{!}{\includegraphics{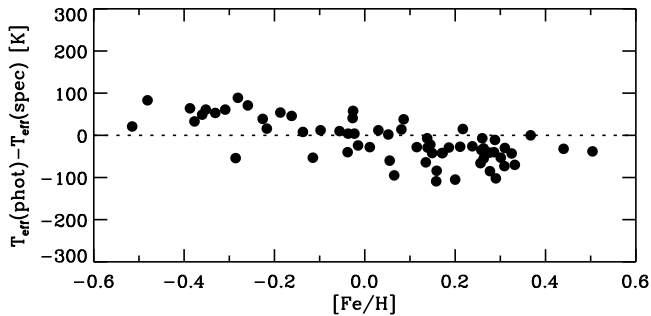}}
\caption{Comparison of photometric temperatures for the HARPS-FEROS 
stars derived in this paper and the spectroscopic
temperatures derived by Sousa et al. (\cite{sousa08}, \cite{sousa11a}, \cite{sousa11b})}
\label{fig:Teff.compare}
\end{figure}

\begin{figure}
\resizebox{\hsize}{!}{\includegraphics{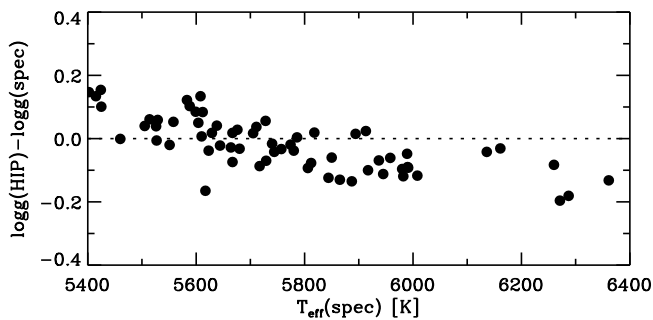}}
\caption{Comparison of surface gravities for the HARPS-FEROS stars
derived via Hipparcos parallaxes in this paper
and the spectroscopic gravities derived by
Sousa et al. (\cite{sousa08}, \cite{sousa11a}, \cite{sousa11b}).}
\label{fig:logg.compare}
\end{figure}

Because of the small error in the Hipparcos parallaxes,
the estimated error in \logg \,(HIP) is about 0.05\,dex.
This small statistical error is supported by the comparison of 
photometric and spectroscopic gravities shown in Fig. \ref{fig:logg.compare};
the rms scatter of $\Delta \logg = \logg \, ({\rm HIP}) - \logg \,({\rm spec})$
in \teff\ intervals of 200\,K is only 0.06\,dex,
but there is a trend of $\Delta \logg$ as a function
of \teff\ ranging from about +0.1 dex at \teff\ = 5400\,K to about $-0.15$\,dex
at the highest temperatures.
This trend is not easily explained; according to  Lind et al. (\cite{lind12})
non-LTE effects on the ionization balance of Fe (used to determine
the spectroscopic gravities) are almost negligible for the effective temperatures,
gravities and metallicities of the present HARPS-FEROS sample of stars.
The same problem is discussed by Bensby et al. (\cite{bensby14}), who note
that  dwarf stars with spectroscopic gravities have a flat distribution
of \logg\ as a function of \teff . In contrast, gravities
derived from Hipparcos parallaxes show an increasing trend of \logg\ when 
\teff\ decreases from 6500\,K to 5000\,K in agreement with predictions from
isochrones.  Hence, it looks like the Hipparcos gravities are more reliable,
but it is puzzling why the spectroscopic gravities have systematic errors.
As discussed by Bensby et al. (\cite{bensby14}), the small departures from LTE 
for the \FeI\ lines predicted with 1D models (Lind et al. \cite{lind12})
cannot explain the problem.
Perhaps one may be able to solve the problem when full 3D, non-LTE calculations
of the ionization balance of Fe becomes available.   

In view of the problems with the spectroscopic gravities, we prefer
to apply the photometric values based on Hipparcos parallaxes. For
consistency we then also adopt the photometric \teff\ values based
on the \by\ and \VK\ colors and the Casagrande et al. (\cite{casagrande10})
calibration. These parameters are given in Table \ref{table:res1}
together with the derived values of \feh\ and \turb , which
agree well with those of Adibekyan et al. (\cite{adibekyan12})
despite of the differences in \teff\ and \logg . Mean differences (this paper
-- Adibekyan) and rms scatters are: $\Delta \, \feh = 0.015 \pm 0.04$
and $\Delta \, \turb = 0.03 \pm 0.09$\,\kmprs . 

\onltab{4}{
\onecolumn
\begin{longtable}{lccrcrrrrrcc}
\caption[ ]{\label{table:res1}Atmospheric parameters and abundances
for the HARPS-FEROS sample of disk stars.} \\
\hline\hline
\noalign{\smallskip}
  ID & \teff & \logg & \feh & \turb & \ch &   \ch & $\oh _{7774}$ & $\oh _{7774}$ & $\oh _{6300}$ &
 Pop.\tablefootmark{a} & $N_{\rm planet}$\tablefootmark{b}  \\
     &  (K)  &       &      &\kmprs & LTE & non-LTE & LTE & non-LTE &     &    &  \\
\noalign{\smallskip}
\hline
\noalign{\smallskip}
\noalign{\smallskip}
\endfirsthead
\caption{continued} \\
\hline\hline
\noalign{\smallskip}
  ID & \teff & \logg & \feh & \turb & \ch &   \ch & $\oh _{7774}$ & $\oh _{7774}$ & $\oh _{6300}$ &
 Pop.\tablefootmark{a} & $N_{\rm planet}$\tablefootmark{b}  \\
     &  (K)  &       &      &\kmprs & LTE & non-LTE & LTE & non-LTE &     &    &  \\
\noalign{\smallskip}
\hline
\noalign{\smallskip}
\noalign{\smallskip}
\endhead
\hline
\endfoot
HD1237 &    5507 &   4.56 &   0.14 &   1.25 &   0.09 &   0.09 &   0.09 &   0.16 &   0.07 &  D & 0  \\
HD4208 &    5688 &   4.53 & $-$0.28 &   1.02 & $-$0.22 & $-$0.22 & $-$0.23 & $-$0.17 &        &  D & 1  \\
HD4307 &    5828 &   4.02 & $-$0.22 &   1.23 & $-$0.24 & $-$0.25 & $-$0.09 & $-$0.15 & $-$0.11 &  D & 0  \\
HD4308 &    5705 &   4.36 & $-$0.35 &   1.00 & $-$0.19 & $-$0.19 & $-$0.04 & $-$0.02 &        &  C & 1  \\
HD14374 &    5466 &   4.58 & $-$0.03 &   0.84 & $-$0.07 & $-$0.07 & $-$0.08 &   0.01 & $-$0.02 &  D & 0  \\
HD16141 &    5722 &   4.10 &   0.16 &   1.13 &   0.09 &   0.09 &   0.18 &   0.12 &        &  D & 1  \\
HD20782 &    5784 &   4.35 & $-$0.06 &   1.02 & $-$0.08 & $-$0.08 & $-$0.04 & $-$0.05 & $-$0.02 &  D & 1  \\
HD23079 &    5988 &   4.38 & $-$0.14 &   1.15 & $-$0.18 & $-$0.18 & $-$0.13 & $-$0.17 & $-$0.09 &  D & 1  \\
HD28185 &    5640 &   4.35 &   0.21 &   0.94 &   0.22 &   0.22 &   0.14 &   0.14 &   0.18 &    D & 1  \\
HD30177 &    5550 &   4.39 &   0.50 &   0.92 &   0.39 &   0.39 &   0.33 &   0.34 &        &    D & 1  \\
HD30306 &    5544 &   4.38 &   0.22 &   0.90 &   0.16 &   0.16 &   0.16 &   0.19 &   0.10 &    D & 0  \\
HD40397 &    5474 &   4.38 & $-$0.12 &   0.96 &   0.05 &   0.05 &   0.11 &   0.16 &   0.05 &    C & 0  \\
HD52265 &    6129 &   4.32 &   0.26 &   1.27 &   0.15 &   0.14 &   0.18 &   0.04 &        &    D & 2  \\
HD65216 &    5658 &   4.52 & $-$0.16 &   0.98 & $-$0.20 & $-$0.19 & $-$0.12 & $-$0.07 & $-$0.08 & D & 2  \\
HD65907 &    5998 &   4.41 & $-$0.33 &   1.08 & $-$0.15 & $-$0.16 &   0.00 & $-$0.05 &   0.03 &    C & 0  \\
HD69830 &    5460 &   4.55 & $-$0.03 &   0.63 & $-$0.04 & $-$0.04 & $-$0.12 & $-$0.03 &   0.00 &    D & 3  \\
HD73256 &    5453 &   4.46 &   0.31 &   1.13 &   0.29 &   0.29 &   0.24 &   0.30 &   0.17 &    D & 1  \\
HD75289 &    6121 &   4.34 &   0.28 &   1.32 &   0.11 &   0.11 &   0.16 &   0.03 &   0.03 &    D & 1  \\
HD77110 &    5738 &   4.39 & $-$0.51 &   0.87 & $-$0.40 & $-$0.40 & $-$0.17 & $-$0.14 & $-$0.07 &  C & 0  \\
HD78538 &    5790 &   4.50 & $-$0.04 &   1.06 & $-$0.17 & $-$0.17 & $-$0.08 & $-$0.07 & $-$0.08 &  D & 0  \\
HD82943 &    5958 &   4.38 &   0.26 &   1.21 &   0.19 &   0.19 &   0.22 &   0.14 &   0.13 &    D & 3  \\
HD89454 &    5699 &   4.53 &   0.19 &   0.94 &   0.06 &   0.06 &   0.12 &   0.14 &   0.04 &    D & 0  \\
HD92788 &    5691 &   4.35 &   0.30 &   1.05 &   0.26 &   0.26 &   0.25 &   0.23 &   0.17 &    D & 2  \\
HD94151 &    5621 &   4.50 &   0.09 &   0.89 &   0.06 &   0.06 &   0.03 &   0.07 &   0.00 &    D & 0  \\
HD96423 &    5689 &   4.39 &   0.14 &   0.96 &   0.08 &   0.08 &   0.11 &   0.11 &   0.08 &    D & 0  \\
HD102365 &    5690 &   4.46 & $-$0.31 &   0.91 & $-$0.19 & $-$0.19 & $-$0.21 & $-$0.17 & $-$0.09 & D & 1  \\
HD108147 &    6218 &   4.39 &   0.17 &   1.35 &   0.05 &   0.05 &   0.14 &   0.00 & $-$0.05 &  D & 1  \\
HD110668 &    5808 &   4.39 &   0.17 &   1.08 &   0.18 &   0.18 &   0.20 &   0.16 &        &    D & 0  \\
HD111232 &    5543 &   4.43 & $-$0.48 &   0.90 & $-$0.20 & $-$0.20 & $-$0.07 & $-$0.01 &        & C & 1  \\
HD114613 &    5700 &   3.90 &   0.14 &   1.33 &   0.11 &   0.10 &   0.16 &   0.07 &   0.07 &   D & 1  \\
HD114729 &    5790 &   4.07 & $-$0.28 &   1.27 & $-$0.21 & $-$0.22 & $-$0.07 & $-$0.12 & $-$0.07 & D & 1  \\
HD114853 &    5744 &   4.46 & $-$0.23 &   0.97 & $-$0.25 & $-$0.25 & $-$0.22 & $-$0.19 &        & D & 0  \\
HD115617 &    5534 &   4.41 & $-$0.01 &   0.91 & $-$0.03 & $-$0.03 & $-$0.07 & $-$0.01 &   0.01 & D & 3  \\
HD117207 &    5632 &   4.34 &   0.26 &   0.97 &   0.20 &   0.20 &   0.18 &   0.19 &        &    D & 1  \\
HD117618 &    5962 &   4.32 &   0.01 &   1.18 &   0.01 &   0.01 &   0.04 & $-$0.03 & $-$0.06 &  D & 2  \\
HD125184 &    5610 &   4.07 &   0.33 &   1.14 &   0.27 &   0.26 &   0.27 &   0.23 &   0.09 &    D & 0  \\
HD125612 &    5872 &   4.45 &   0.28 &   1.13 &   0.10 &   0.10 &   0.17 &   0.12 &   0.18 &    D & 3  \\
HD126525 &    5650 &   4.41 & $-$0.10 &   0.92 & $-$0.06 & $-$0.06 & $-$0.09 & $-$0.06 &   0.00 &  D & 1  \\
HD128674 &    5600 &   4.48 & $-$0.36 &   0.82 & $-$0.32 & $-$0.32 & $-$0.30 & $-$0.22 & $-$0.23 & D & 0  \\
HD134664 &    5805 &   4.39 &   0.05 &   1.11 & $-$0.01 & $-$0.01 &   0.02 &   0.00 &   0.00 &    D & 0  \\
HD134987 &    5710 &   4.28 &   0.31 &   1.08 &   0.30 &   0.30 &   0.27 &   0.23 &   0.16 &    D & 2  \\
HD136352 &    5728 &   4.36 & $-$0.39 &   1.05 & $-$0.18 & $-$0.18 & $-$0.08 & $-$0.07 &   0.00 & C & 3  \\
HD140901 &    5582 &   4.47 &   0.12 &   0.95 &   0.08 &   0.08 &   0.08 &   0.12 & $-$0.03 &    D & 0  \\
HD145666 &    5918 &   4.47 & $-$0.04 &   1.09 & $-$0.10 & $-$0.10 & $-$0.01 & $-$0.03 & $-$0.06 & D & 0  \\
HD146233 &    5820 &   4.47 &   0.05 &   1.01 & $-$0.02 & $-$0.02 & $-$0.01 & $-$0.01 &   0.03 & D & 0 \\
HD157347 &    5688 &   4.41 &   0.03 &   0.96 & $-$0.03 & $-$0.03 &   0.00 &   0.01 &   0.04 &  D & 0  \\
HD160691 &    5737 &   4.23 &   0.32 &   1.11 &   0.28 &   0.28 &   0.28 &   0.22 &   0.21 &    D & 4  \\
HD168443 &    5522 &   4.06 &   0.07 &   1.02 &   0.17 &   0.16 &   0.23 &   0.22 &   0.14 &    C & 2  \\
HD169830 &    6319 &   4.08 &   0.15 &   1.58 &   0.11 &   0.10 &   0.28 &   0.01 &   0.11 &    D & 2  \\
HD179949 &    6182 &   4.36 &   0.20 &   1.34 &   0.12 &   0.12 &   0.21 &   0.06 &        &    D & 1  \\
HD183263 &    5889 &   4.29 &   0.29 &   1.23 &   0.25 &   0.25 &   0.22 &   0.14 &        &    D & 2  \\
HD190248 &    5572 &   4.31 &   0.44 &   0.94 &   0.38 &   0.38 &   0.34 &   0.34 &   0.22 &    D & 0  \\
HD196050 &    5862 &   4.22 &   0.26 &   1.16 &   0.24 &   0.23 &   0.23 &   0.14 &   0.14 &    D & 1  \\
HD196761 &    5486 &   4.56 & $-$0.26 &   0.75 & $-$0.22 & $-$0.21 & $-$0.32 & $-$0.22 &   &    D & 0  \\
HD202206 &    5746 &   4.44 &   0.29 &   1.11 &   0.15 &   0.15 &   0.09 &   0.08 &   0.05 &    D & 2  \\
HD203608 &    6180 &   4.35 & $-$0.66 &   1.30 & $-$0.57 & $-$0.58 & $-$0.41 & $-$0.45 & $-$0.42 & D & 0  \\
HD206172 &    5662 &   4.62 & $-$0.19 &   0.84 & $-$0.21 & $-$0.21 & $-$0.11 & $-$0.05 &   &    D & 0  \\
HD207129 &    5941 &   4.42 & $-$0.02 &   1.06 & $-$0.11 & $-$0.12 & $-$0.06 & $-$0.09 &   0.00 & D & 0  \\
HD210277 &    5479 &   4.34 &   0.24 &   0.84 &   0.28 &   0.28 &   0.26 &   0.29 &   0.17 &    C & 1  \\
HD212301 &    6162 &   4.35 &   0.16 &   1.35 &   0.11 &   0.10 &   0.17 &   0.03 &        &    D & 1  \\
HD213240 &    5918 &   4.15 &   0.13 &   1.25 &   0.10 &   0.09 &   0.16 &   0.05 &   0.04 &    D & 1  \\
HD216435 &    5942 &   4.08 &   0.25 &   1.36 &   0.20 &   0.19 &   0.26 &   0.11 &   0.07 &    D & 1  \\
HD216437 &    5802 &   4.17 &   0.28 &   1.14 &   0.24 &   0.24 &   0.25 &   0.17 &   0.17 &    D & 1  \\
HD216770 &    5424 &   4.53 &   0.37 &   0.79 &   0.29 &   0.30 &   0.18 &   0.26 &   0.17 &    C & 1  \\
HD216777 &    5656 &   4.47 & $-$0.38 &   0.89 & $-$0.35 & $-$0.35 & $-$0.28 & $-$0.22 &   &    D & 0  \\
HD222669 &    5908 &   4.48 &   0.08 &   1.01 &   0.00 &   0.00 &   0.02 & $-$0.01 &  0.02 &    D & 0  \\
\noalign{\smallskip}
\hline

\end{longtable}

\tablefoot{
\tablefoottext{a}{Population classification: C, thick-disk; D, thin-disk.} \\ 
\tablefoottext{b} {Number of planets detected, ({\tt http://exoplanets.org}, April 2014).}}

}

\subsection{The UVES-FIES sample}

This sample consists of halo and thick-disk stars 
with distances up to 350\,pc, which means that
interstellar reddening may affect the effective temperature determined
from color indices and that in most cases the Hipparcos parallax
has a too large error to be used to determine the surface gravity. 
Hence, it is
not possible to determine precise photometric values of \teff\ and \logg . 
Instead, we have followed
the method applied by Nissen \& Schuster (\cite{nissen10}, \cite{nissen11}), 
who first determined \teff\  and \logg\ for two nearby, unreddened
thick-disk stars,
\object{HD\,22879} and \object{HD\,76932}, with the photometric method
described in the previous section.  These standard stars are analyzed 
relative to the Sun using lines having EWs less than about 100\,m\AA\ 
in the solar spectrum. An inverted abundance analysis then yields $gf$-values
for the full set of lines given in Nissen \& Schuster (\cite{nissen11}, Table 3).
Using these $gf$-values, a model atmosphere analysis makes it possible to 
determine spectroscopic values of \teff\ and \logg\ as well as 
high-precision abundances relative to the standard stars.

In Nissen \& Schuster (\cite{nissen10}, \cite{nissen11}), the calibrations of
Ram\'{\i}rez \& Mel\'{e}ndez (\cite{ramirez05}) were used to determine \teff\
of the standard stars. 
The more accurate calibrations by Casagrande et al.
(\cite{casagrande10}) show, however, a systematic offset of
about +100\,K in \teff\ relative to the Ram\'{\i}rez \& Mel\'endez values.
We have, therefore, increased the \teff\ values of the standard stars
by 100\,K, and repeated the spectroscopic analysis of the other stars, i.e.,
determined \teff\ from the excitation balance of weak ($EW < 50$\,m\AA )
\FeI\ lines, \logg\ from the \FeI /\FeII\ ionization balance,
and \turb\  by requesting that the derived \feh\
has no systematic dependence on EW for \FeI\ lines.
The updated parameters are given in Table \ref{table:res2} for 85 stars for which
C and/or O abundances have been determined\footnote{The C and $s$-process rich
CH-subgiant, \object{G\,24-25} (Liu et al. \cite{liushu12}) 
is not included.}. Relative to the parameter 
values given in Nissen \& Schuster (\cite{nissen10}, \cite{nissen11}), there are 
small changes in \logg\ ranging from +0.03\,dex for the 
standard stars to about +0.08\,dex
for the coolest stars. \feh\ is decreased by approximately 
0.02\,dex for all stars, whereas
\alphafe\ is practically the same, i.e., changes are within  $\pm 0.005$\,dex.

\onltab{5}{
\onecolumn
\begin{longtable}{lccccccccccc}
\caption{\label{table:res2}Atmospheric parameters and abundances for
the UVES-FIES sample of halo and thick-disk stars.} \\
\hline\hline
\noalign{\smallskip}
  ID & \teff & \logg & \feh & \turb & \ch &   \ch & $\oh _{7774}$ & $\oh _{7774}$ & $\oh _{6300}$ 
& Pop.\tablefootmark{a} & Bin.\tablefootmark{b}  \\
     &  (K)  &       &      &\kmprs & LTE & non-LTE & LTE & non-LTE &     &  &  \\
\noalign{\smallskip}
\hline
\noalign{\smallskip}
\noalign{\smallskip}
\endfirsthead
\caption{continued} \\
\hline\hline
\noalign{\smallskip}
  ID & \teff & \logg & \feh & \turb & \ch &   \ch & $\oh _{7774}$ & $\oh _{7774}$ & $\oh _{6300}$ 
& Pop.\tablefootmark{a} & Bin.\tablefootmark{b}  \\
     &  (K)  &       &      &\kmprs & LTE & non-LTE & LTE & non-LTE &     &  &  \\
\noalign{\smallskip}
\hline
\noalign{\smallskip}
\noalign{\smallskip}
\endhead
\hline
\endfoot
 BD$-$21 3420 &    5909 &   4.30 &  $-$1.14 &   1.12 &  $-$0.91 &  $-$0.91 &  $-$0.50 &  $-$0.49 &  $-$0.52 &    C & \\
 CD$-$33 3337 &    6112 &   3.86 &  $-$1.37 &   1.56 &  $-$1.19 &  $-$1.21 &  $-$0.81 &  $-$0.86 &        &    C & \\
 CD$-$43 6810 &    6059 &   4.32 &  $-$0.44 &   1.24 &  $-$0.24 &  $-$0.24 &   0.00 &  $-$0.08 &   0.00 &    A & \\
 CD$-$45 3283 &    5685 &   4.61 &  $-$0.93 &   0.95 &  $-$1.02 &  $-$1.02 &  $-$0.60 &  $-$0.50 &        &    B & \\
 CD$-$51 4628 &    6296 &   4.29 &  $-$1.32 &   1.31 &  $-$1.40 &  $-$1.41 &  $-$0.98 &  $-$0.98 &        &    B & \\
 CD$-$57 1633 &    5981 &   4.29 &  $-$0.91 &   1.08 &  $-$0.96 &  $-$0.97 &  $-$0.64 &  $-$0.62 &        &    B & \\
 CD$-$61 0282 &    5869 &   4.34 &  $-$1.25 &   1.19 &  $-$1.29 &  $-$1.29 &  $-$0.75 &  $-$0.70 &        &    B & \\
 G05$-$19 &    5770 &   4.28 &  $-$1.19 &   1.17 &  $-$1.25 &  $-$1.25 &  $-$0.78 &  $-$0.71 &        &    B & \\
 G05$-$36 &    6139 &   4.22 &  $-$1.25 &   1.29 &  $-$1.12 &  $-$1.12 &  $-$0.67 &  $-$0.69 &        &    A & \\
 G05$-$40 &    5892 &   4.20 &  $-$0.83 &   1.12 &  $-$0.62 &  $-$0.63 &  $-$0.27 &  $-$0.29 &        &    A & \\
 G15$-$23 &    5373 &   4.63 &  $-$1.12 &   0.90 &  $-$0.87 &  $-$0.87 &  $-$0.52 &  $-$0.38 &        &    A & \\
 G18$-$28 &    5443 &   4.49 &  $-$0.85 &   0.88 &  $-$0.59 &  $-$0.59 &  $-$0.31 &  $-$0.20 &        &    A & SB1 \\
 G18$-$39 &    6175 &   4.21 &  $-$1.41 &   1.37 &  $-$1.32 &  $-$1.33 &  $-$0.88 &  $-$0.88 &        &    A & \\
 G20$-$15 &    6162 &   4.32 &  $-$1.50 &   1.50 &  $-$1.44 &  $-$1.45 &  $-$1.07 &  $-$1.04 &        &    B & \\
 G21$-$22 &    6021 &   4.27 &  $-$1.10 &   1.30 &  $-$1.19 &  $-$1.19 &        &        &        &    B & \\
 G24$-$13 &    5764 &   4.38 &  $-$0.73 &   0.86 &  $-$0.59 &  $-$0.59 &  $-$0.14 &  $-$0.12 &        &    A & \\
 G31$-$55 &    5731 &   4.35 &  $-$1.12 &   1.26 &  $-$0.92 &  $-$0.92 &  $-$0.58 &  $-$0.51 &        &    A & \\
 G46$-$31 &    6017 &   4.29 &  $-$0.83 &   1.30 &  $-$0.76 &  $-$0.77 &  $-$0.50 &  $-$0.51 &  $-$0.41 &    B & SB1 \\
 G49$-$19 &    5863 &   4.32 &  $-$0.55 &   1.12 &  $-$0.37 &  $-$0.37 &        &        &        &    A & SB1  \\
 G53$-$41 &    5975 &   4.29 &  $-$1.21 &   1.20 &  $-$1.38 &  $-$1.39 &  $-$1.06 &  $-$1.00 &        &    B & \\
 G56$-$30 &    5935 &   4.29 &  $-$0.90 &   1.22 &  $-$1.02 &  $-$1.02 &        &        &        &    B & \\
 G56$-$36 &    6067 &   4.33 &  $-$0.94 &   1.33 &  $-$0.87 &  $-$0.88 &  $-$0.51 &  $-$0.52 &        &    B & \\
 G57$-$07 &    5755 &   4.33 &  $-$0.48 &   0.99 &  $-$0.33 &  $-$0.33 &        &        &        &    A & \\
 G63$-$26 &    6175 &   4.17 &  $-$1.58 &   1.65 &  $-$1.36 &  $-$1.37 &        &        &        &    A & \\
 G66$-$22 &    5297 &   4.46 &  $-$0.88 &   0.78 &  $-$0.98 &  $-$0.98 &  $-$0.59 &  $-$0.46 &        &    B & \\
 G74$-$32 &    5864 &   4.41 &  $-$0.74 &   1.04 &  $-$0.54 &  $-$0.54 &        &        &        &    A & \\
 G75$-$31 &    6135 &   4.02 &  $-$1.04 &   1.28 &  $-$1.17 &  $-$1.18 &  $-$0.72 &  $-$0.75 &        &    B & \\
 G81$-$02 &    5967 &   4.24 &  $-$0.69 &   1.21 &  $-$0.57 &  $-$0.57 &        &        &        &    A & \\
 G82$-$05 &    5338 &   4.51 &  $-$0.78 &   0.80 &  $-$0.69 &  $-$0.69 &  $-$0.51 &  $-$0.38 &        &    B & \\
 G85$-$13 &    5709 &   4.46 &  $-$0.60 &   0.87 &  $-$0.46 &  $-$0.46 &  $-$0.15 &  $-$0.11 &        &    A & \\
 G87$-$13 &    6217 &   4.11 &  $-$1.10 &   1.42 &  $-$1.34 &  $-$1.35 &        &        &        &    B & \\
 G96$-$20 &    6445 &   4.46 &  $-$0.90 &   1.42 &  $-$0.62 &  $-$0.63 &        &        &        &    A & \\
 G98$-$53 &    5954 &   4.26 &  $-$0.89 &   1.20 &  $-$0.83 &  $-$0.84 &        &        &        &    B & \\
 G99$-$21 &    5559 &   4.46 &  $-$0.68 &   0.79 &  $-$0.41 &  $-$0.41 &  $-$0.16 &  $-$0.09 &        &    A & \\
 G112$-$43 &    6209 &   4.02 &  $-$1.27 &   1.17 &  $-$1.23 &  $-$1.24 &  $-$0.88 &  $-$0.91 &        &    B & \\
 G112$-$44 &    5936 &   4.28 &  $-$1.31 &   1.10 &  $-$1.21 &  $-$1.21 &        &        &        &    B & \\
 G114$-$42 &    5721 &   4.40 &  $-$1.12 &   1.19 &  $-$1.10 &  $-$1.11 &        &        &        &    B & \\
 G119$-$64 &    6333 &   4.14 &  $-$1.50 &   1.40 &  $-$1.53 &  $-$1.54 &  $-$1.08 &  $-$1.09 &        &    B & \\
 G121$-$12 &    6041 &   4.25 &  $-$0.94 &   1.26 &  $-$1.09 &  $-$1.10 &        &        &        &    B & \\
 G127$-$26 &    5886 &   4.20 &  $-$0.53 &   1.11 &  $-$0.37 &  $-$0.38 &        &        &        &    A & \\
 G150$-$40 &    6080 &   4.11 &  $-$0.82 &   1.31 &  $-$1.07 &  $-$1.08 &  $-$0.63 &  $-$0.66 &        &    B & \\
 G159$-$50 &    5713 &   4.44 &  $-$0.94 &   1.03 &  $-$0.70 &  $-$0.70 &  $-$0.48 &  $-$0.41 &        &    A & \\
 G161$-$73 &    6108 &   3.99 &  $-$1.01 &   1.26 &  $-$1.18 &  $-$1.19 &        &        &        &    B & \\
 G170$-$56 &    6112 &   4.11 &  $-$0.94 &   1.39 &  $-$0.96 &  $-$0.97 &  $-$0.62 &  $-$0.65 &        &    B & \\
 G176$-$53 &    5615 &   4.52 &  $-$1.36 &   0.90 &        &        &  $-$0.85 &  $-$0.73 &        &    B & \\
 G180$-$24 &    6137 &   4.20 &  $-$1.41 &   1.45 &  $-$1.17 &  $-$1.18 &  $-$0.82 &  $-$0.82 &        &    A & \\
 G187$-$18 &    5691 &   4.46 &  $-$0.68 &   1.05 &  $-$0.44 &  $-$0.44 &        &        &        &    A & \\
 G188$-$22 &    6116 &   4.20 &  $-$1.33 &   1.42 &  $-$1.18 &  $-$1.19 &  $-$0.68 &  $-$0.70 &        &    A & \\
HD3567 &    6180 &   4.01 &  $-$1.17 &   1.40 &  $-$1.26 &  $-$1.28 &  $-$0.83 &  $-$0.86 &        &    B & \\
HD17820 &    5873 &   4.28 &  $-$0.68 &   1.27 &  $-$0.46 &  $-$0.47 &  $-$0.18 &  $-$0.20 &  $-$0.21 &    C & \\
HD22879 &    5859 &   4.29 &  $-$0.86 &   1.20 &  $-$0.64 &  $-$0.64 &  $-$0.29 &  $-$0.29 &        &    C & \\
HD25704 &    5974 &   4.30 &  $-$0.86 &   1.33 &  $-$0.67 &  $-$0.68 &  $-$0.44 &  $-$0.44 &  $-$0.40 &    C & D \\
HD51754 &    5857 &   4.35 &  $-$0.58 &   1.30 &  $-$0.39 &  $-$0.39 &  $-$0.16 &  $-$0.17 &  $-$0.19 &    A & \\
HD59392 &    6137 &   3.88 &  $-$1.62 &   1.73 &  $-$1.61 &  $-$1.62 &  $-$1.12 &  $-$1.12 &        &    B & \\
HD76932 &    5977 &   4.17 &  $-$0.87 &   1.30 &  $-$0.73 &  $-$0.73 &  $-$0.37 &  $-$0.40 &  $-$0.30 &    C & \\
HD97320 &    6136 &   4.20 &  $-$1.18 &   1.46 &  $-$0.96 &  $-$0.97 &  $-$0.66 &  $-$0.68 &        &    C & \\
HD103723 &    6050 &   4.20 &  $-$0.81 &   1.11 &  $-$0.86 &  $-$0.86 &  $-$0.54 &  $-$0.56 &        &    B & \\
HD105004 &    5852 &   4.35 &  $-$0.83 &   1.09 &  $-$0.73 &  $-$0.73 &  $-$0.56 &  $-$0.52 &        &    B & \\
HD106516 &    6327 &   4.43 &  $-$0.69 &   1.18 &  $-$0.51 &  $-$0.51 &  $-$0.11 &  $-$0.22 &        &    C & SB1 \\
HD111980 &    5878 &   3.98 &  $-$1.09 &   1.39 &  $-$0.92 &  $-$0.93 &  $-$0.43 &  $-$0.46 &  $-$0.53 &  A & SB1 \\ 
HD113679 &    5761 &   4.05 &  $-$0.66 &   1.37 &  $-$0.53 &  $-$0.53 &  $-$0.17 &  $-$0.19 &  $-$0.24 &    A & \\
HD114762A&    5956 &   4.24 &  $-$0.72 &   1.37 &  $-$0.53 &  $-$0.54 &  $-$0.28 &  $-$0.31 &        &    C & SB1 \\
HD120559 &    5486 &   4.58 &  $-$0.91 &   1.05 &  $-$0.61 &  $-$0.60 &  $-$0.39 &  $-$0.26 &  $-$0.25 &    C & \\
HD121004 &    5755 &   4.43 &  $-$0.71 &   1.16 &  $-$0.52 &  $-$0.52 &  $-$0.26 &  $-$0.22 &  $-$0.20 &    A & \\
HD126681 &    5594 &   4.50 &  $-$1.20 &   1.08 &  $-$0.98 &  $-$0.98 &  $-$0.61 &  $-$0.50 &        &    C & \\
HD132475 &    5750 &   3.77 &  $-$1.51 &   1.37 &  $-$1.35 &  $-$1.36 &  $-$0.82 &  $-$0.79 &        &    A & \\
HD148816 &    5923 &   4.17 &  $-$0.74 &   1.33 &  $-$0.53 &  $-$0.53 &  $-$0.24 &  $-$0.28 &        &    A & \\
HD159482 &    5829 &   4.37 &  $-$0.74 &   1.21 &  $-$0.54 &  $-$0.55 &  $-$0.20 &  $-$0.19 &        &    A & D \\
HD160693 &    5809 &   4.35 &  $-$0.48 &   1.02 &  $-$0.37 &  $-$0.37 &  $-$0.09 &  $-$0.10 &        &    A & \\
HD163810 &    5592 &   4.61 &  $-$1.22 &   1.17 &  $-$1.17 &  $-$1.17 &  $-$0.79 &  $-$0.67 &        &    B & D \\
HD175179 &    5804 &   4.40 &  $-$0.66 &   1.08 &  $-$0.42 &  $-$0.42 &  $-$0.09 &  $-$0.09 &  $-$0.14 &    C & \\
HD179626 &    5957 &   4.16 &  $-$1.06 &   1.47 &  $-$0.87 &  $-$0.87 &  $-$0.47 &  $-$0.48 &        &    A & \\
HD189558 &    5707 &   3.83 &  $-$1.14 &   1.29 &  $-$1.05 &  $-$1.06 &  $-$0.51 &  $-$0.51 &        &    C & \\
HD193901 &    5745 &   4.42 &  $-$1.11 &   1.12 &  $-$1.05 &  $-$1.05 &  $-$0.71 &  $-$0.63 &        &    B & \\
HD194598 &    6053 &   4.33 &  $-$1.11 &   1.30 &  $-$1.13 &  $-$1.14 &  $-$0.72 &  $-$0.70 &        &    B & \\
HD199289 &    5915 &   4.30 &  $-$1.05 &   1.21 &  $-$0.87 &  $-$0.88 &  $-$0.50 &  $-$0.49 &  $-$0.47 &    C & \\
HD205650 &    5793 &   4.35 &  $-$1.19 &   1.17 &  $-$0.94 &  $-$0.94 &  $-$0.60 &  $-$0.54 &        &    C  & \\
HD219617 &    5983 &   4.28 &  $-$1.46 &   1.42 &  $-$1.53 &  $-$1.53 &  $-$0.95 &  $-$0.91 &        &    B & D \\
HD222766 &    5423 &   4.38 &  $-$0.70 &   0.75 &  $-$0.41 &  $-$0.41 &  $-$0.16 &  $-$0.07 &        &    A & \\
HD230409 &    5386 &   4.61 &  $-$0.87 &   1.01 &  $-$0.71 &  $-$0.70 &  $-$0.40 &  $-$0.27 &        &    A & \\
HD233511 &    6125 &   4.21 &  $-$1.58 &   1.20 &        &        &  $-$1.02 &  $-$0.99 &        &    A & \\
HD237822 &    5675 &   4.41 &  $-$0.47 &   0.99 &  $-$0.23 &  $-$0.23 &        &        &        &    A & \\
HD241253 &    5940 &   4.34 &  $-$1.11 &   1.17 &  $-$0.92 &  $-$0.93 &  $-$0.52 &  $-$0.51 &        &    C & \\
HD250792A&    5572 &   4.50 &  $-$1.03 &   0.98 &        &        &  $-$0.62 &  $-$0.50 &        &    B & D \\
HD284248 &    6271 &   4.21 &  $-$1.59 &   1.51 &  $-$1.52 &  $-$1.53 &  $-$1.17 &  $-$1.15 &        &    B & \\
\noalign{\smallskip}
\hline

\end{longtable}

\tablefoot{
\tablefoottext{a}{Population classification: A, high-alpha halo; B, low-alpha halo;
C, thick-disk.} \\
\tablefoottext{b}{Note on binarity: SB1, single-lined spectroscopic binary according to the SIMBAD database;
D, double star according to the HIPPARCOS and TYCHO catalogues (Perryman et al. \cite{perryman97}).} }

}

As a check of the spectroscopic values of
\teff , we have compared with IRFM values from Casagrande et al.
(\cite{casagrande10}) for a subsample of 41 stars in common.
As seen from Fig. \ref{fig:Teff.compareIRFM}, there is no
significant trend of $\Delta \teff \, = \, \teff (\rm {spec}) - \teff (\rm {IRFM})$
as a function of \feh , nor is there any trend of $\Delta \teff$ with \teff .
The mean value of  $\Delta \teff$ is $-10$\,K with an rms scatter of $\pm 70$\,K.
The largest contribution to this scatter arises from the error in \teff (\rm {IRFM}),
which is on the order of $\pm 60$\,K. These numbers suggest that the error in 
\teff (spec) is on the order of $\pm 35$\,K.

\begin{figure}
\resizebox{\hsize}{!}{\includegraphics{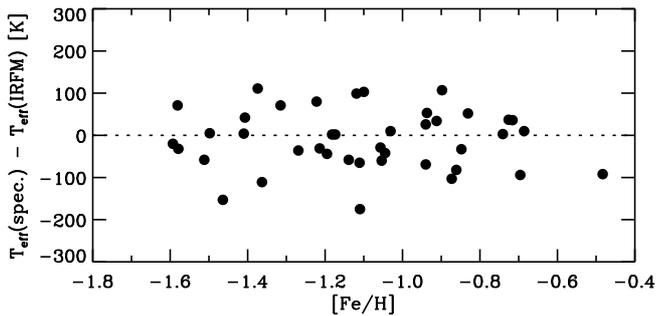}}
\caption{Comparison of spectroscopic temperatures derived
in this paper for the UVES-FIES sample and 
temperatures determined by Casagrande et al. (\cite{casagrande10}) with
the IRFM method.}
\label{fig:Teff.compareIRFM}
\end{figure}

\begin{figure}
\resizebox{\hsize}{!}{\includegraphics{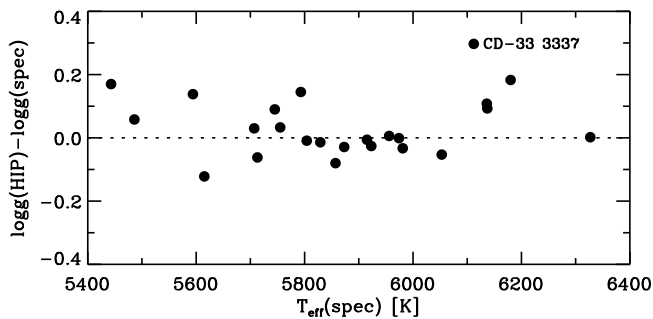}}
\caption{Comparison of photometric gravities derived via Hipparcos parallaxes
and spectroscopic gravities derived from \FeI\ and \FeII\ lines for a subsample of
the UVES-FIES stars having parallax errors less than 10\% .}
\label{fig:logg.compare.halorich}
\end{figure}

The spectroscopic gravities have been tested by comparing with 
gravities based on Hipparcos parallaxes for 
24 stars\footnote{The two standard stars are not included in this
comparison, because their spectroscopic and photometric gravities agree by
definition.}, which have no detectable interstellar NaD lines (indicating
that interstellar reddening is negligible) and parallaxes with an error of less
than 10\%. As seen from Fig. \ref{fig:logg.compare.halorich}, there is
a satisfactory agreement between the two sets of gravities although one
star, \object{CD\,$-33 \, 3337$}, shows a large deviation. If this outlier
is excluded, the mean of $\Delta \logg = \logg$\,(HIP) $- \logg$\,(spec)
is 0.03\,dex with an rms scatter of $\pm 0.085$\,dex. With an estimated
average error of $\pm 0.06$\,dex for \logg \,(HIP), the error of \logg \,(spec)
becomes $\pm 0.06$\,dex. If \object{CD\,$-33 \, 3337$} is included,
the scatter of  $\Delta \logg$ raises to $\pm 0.10$\,dex and the error of
\logg \,(spec) to $\pm 0.08$\,dex.

Interestingly, there is no indication of a trend of $\Delta \logg$ as a
function of \teff\ like the trend
seen in Fig. \ref{fig:Teff.compare} for the more metal-rich
disk stars. Thus, it seems that we can use spectroscopic values of \logg\
for the FIES -- UVES sample of halo and thick-disk stars without worrying about
systematic errors as a function of \teff .

\section{Carbon and Oxygen abundances}
\label{sect:COabundances}
The derived LTE and non-LTE values of \ch\ and \oh\ are given in
Tables \ref{table:res1} and \ref{table:res2}. In the
case of \oh\ we list both the value derived from the \OI\ triplet and the
value based on the \oI \, $\lambda 6300$ line, if measured. Most
of the disk stars have $\oh _{6300}$ available, whereas the 
\oI\ line is too weak for the majority of halo stars to 
be detectable.

\subsection{Statistical errors}
\label{sect:stat.errors}
Statistical errors of the various abundance ratios have been determined
for two representative stars: a disk star, \object{HD\,196050}, with $\feh = +0.26$ and 
a metal-poor star from the UVES-FIES sample, \object{HD\,199289}, with $\feh = -1.05$.
As seen from Tables \ref{table:error1} and \ref{table:error2},
$\oh _{6300}$ derived from the \oI\ $\lambda 6300$ line
is only slightly affected by the uncertainty in \teff ,
whereas $\oh_{7774}$ derived from the \OI\ triplet
lines is sensitive to the error in \teff . \feh , \ch , and \oh\ are all significantly
affected by the error in \logg , but this dependence cancels out for \cfe , \ofe ,
and \co . Furthermore, we note that the errors in \feh\ and \turb\ are of minor
importance for the derived abundance ratios.

The errors of \feh , \ch , and $\oh_{7774}$
arising from the uncertainty of the measured equivalent widths
have been estimated from the line-to-line scatter of the derived 
abundances. In the case of $\oh _{6300}$, the EW-error corresponds
to an uncertainty of the equivalent width of the $\lambda 6300$ \oI\ 
line of $\pm 0.4$\,m\AA\ as estimated in Sect. \ref{sect:HARPS-FEROS}.

The total errors listed in the last rows of 
Tables \ref{table:error1} and \ref{table:error2} are 
calculated by adding the individual errors in
quadrature. The small estimated error of \cfe\ ($\pm 0.028$\,dex)
in Table \ref{table:error1} is due to the extremely high S/N
of the HARPS spectra, and the small sensitivity of \cfe\ to
uncertainties in the atmospheric parameters. For the UVES-FIES sample
the error of \cfe\ is larger, because of the weakness of the 
\CI\ lines, especially in spectra of metal-poor halo stars belonging
to the low-alpha population. 
In some of the figures in Sect. \ref{sect:discussion}, data will be shown
with individual error bars.

\begin{table*}
\caption[ ]{Errors in abundance ratios of the metal-rich star \object{HD\,196050}\tablefootmark{a}
caused by errors in model atmosphere parameters and equivalent widths.}
\label{table:error1}
\setlength{\tabcolsep}{0.20cm}
\begin{tabular}{lcccccccc}
\noalign{\smallskip}
\hline\hline
\noalign{\smallskip}
     & $\sigma \feh $ & $\sigma \ch $ & $\sigma \oh _{6300}$\tablefootmark{b}
 & $\sigma \oh _{7774}$\tablefootmark{c}  
     & $\sigma \cfe $ & $\sigma \ofe _{6300}$\tablefootmark{b} 
& $\sigma \ofe _{7774}$\tablefootmark{c} & $\sigma \co _{7774}$\tablefootmark{c} \\
\noalign{\smallskip}
\hline
\noalign{\smallskip}
$\sigma (\teff ) = \pm 30$\,K              &  $\mp 0.011$  & $\mp 0.020$ & $\pm 0.002$ 
 & $\mp 0.035$ & $\mp 0.009$ & $\pm 0.013$ & $\mp 0.024$ & $\pm 0.015$  \\
$\sigma (\logg ) = \pm 0.05$\,dex          &  $\pm 0.020$  & $\pm 0.017$ & $\pm 0.023$   
 & $\pm 0.021$ & $\mp 0.003$ & $\pm 0.003$ & $\pm 0.001$ & $\mp 0.004$  \\
$\sigma \feh  = \pm 0.03$\,dex             &  $\pm 0.008$  & $\pm 0.000$ & $\pm 0.010$
 & $\pm 0.001$ & $\mp 0.008$ & $\pm 0.002$ & $\mp 0.007$ & $\mp 0.001$   \\
$\sigma (\turb ) = \pm 0.06$\,\kmprs       &  $\mp 0.014$  & $\mp 0.003$ & $\pm 0.000$  
 & $\mp 0.005$ & $\pm 0.011$ & $\pm 0.014$ & $\pm 0.009$ & $\pm 0.002  $ \\
$\sigma (EW) $                             &  $\pm 0.010$  & $\pm 0.020$ & $\pm 0.040$  
 & $\pm 0.025$ & $\pm 0.022$ & $\pm 0.041$ & $\pm 0.027$ & $\pm 0.032  $ \\
\noalign{\smallskip}
\hline
\noalign{\smallskip}
$\sigma$\,(total)                          &  $\pm 0.030$  & $\pm 0.033$ & $\pm 0.047$  
 & $\pm 0.048$ & $\pm 0.028$ & $\pm 0.045$ & $\pm 0.038$ & $\pm 0.036  $ \\
\noalign{\smallskip}
\hline
\end{tabular}

\tablefoot{
\tablefoottext{a}{(\teff , \logg , \feh ) = (5852\,K, 4.22, +0.26)}
\tablefoottext{b}{The oxygen abundance is based on the $\oI \, \lambda 6300$ line.}
\tablefoottext{c}{The oxygen abundance is based on the $\OI \, \lambda 7774$ triplet.}}

\end{table*}

\begin{table*}
\caption[ ]{Errors in abundance ratios of the metal-poor star \object{HD\,199289}\tablefootmark{a}
caused by errors in model atmosphere parameters and equivalent widths.}
\label{table:error2}
\setlength{\tabcolsep}{0.20cm}
\begin{tabular}{lcccccccc}
\noalign{\smallskip}
\hline\hline
\noalign{\smallskip}
     & $\sigma \feh $ & $\sigma \ch $ & $\sigma \oh _{6300}$\tablefootmark{b}
 & $\sigma \oh _{7774}$\tablefootmark{c}
     & $\sigma \cfe $ & $\sigma \ofe _{6300}$\tablefootmark{b}
& $\sigma \ofe _{7774}$\tablefootmark{c} & $\sigma \co _{7774}$\tablefootmark{c} \\
\noalign{\smallskip}
\hline
\noalign{\smallskip}
$\sigma (\teff ) = \pm 35$\,K              &  $\mp 0.002$  & $\mp 0.018$ & $\pm 0.012$
 & $\mp 0.034$ & $\mp 0.016$ & $\pm 0.014$ & $\mp 0.032$ & $\pm 0.016$  \\
$\sigma (\logg ) = \pm 0.06$\,dex          &  $\pm 0.022$  & $\pm 0.024$ & $\pm 0.024$
 & $\pm 0.023$ & $\pm 0.002$ & $\pm 0.002$ & $\pm 0.001$ & $\pm 0.001$  \\
$\sigma \feh  = \pm 0.03$\,dex             &  $\pm 0.002$  & $\mp 0.002$ & $\pm 0.004$
 & $\mp 0.001$ & $\mp 0.004$ & $\pm 0.002$ & $\mp 0.003$ & $\mp 0.001$   \\
$\sigma (\turb ) = \pm 0.06$\,\kmprs       &  $\mp 0.008$  & $\pm 0.000$ & $\pm 0.000$
 & $\mp 0.003$ & $\pm 0.008$ & $\pm 0.008$ & $\pm 0.005$ & $\pm 0.003  $ \\
$\sigma (EW) $                             &  $\pm 0.015$  & $\pm 0.045$ & $\pm 0.050$
 & $\pm 0.025$ & $\pm 0.047$ & $\pm 0.052$ & $\pm 0.029$ & $\pm 0.051  $ \\
\noalign{\smallskip}
\hline
\noalign{\smallskip}
$\sigma$\,(total)                          &  $\pm 0.028$  & $\pm 0.054$ & $\pm 0.057$  
 & $\pm 0.048$ & $\pm 0.050$ & $\pm 0.055$ & $\pm 0.044$ & $\pm 0.054  $ \\
\noalign{\smallskip}
\hline
\end{tabular}

\tablefoot{
\tablefoottext{a}{(\teff , \logg , \feh ) = (5915\,K, 4.30, $-1.05$)}
\tablefoottext{b}{The oxygen abundance is based on the $\oI \, \lambda 6300$ line.}
\tablefoottext{c}{The oxygen abundance is based on the $\OI \, \lambda 7774$ triplet.}}

\end{table*}

\subsection{Systematic errors; the $\oh_{7774}$ - $\oh_{6300}$ discrepancy.}
\label{sect:sys.errors}
In addition to statistical errors, we have to consider possible systematic
errors of the abundances, especially in the case of oxygen abundances derived
from the $\OI \, \lambda 7774$ lines, because of the large non-LTE corrections
and high sensitivity to \teff . In order to reveal
any problems, we have therefore compared oxygen abundances derived from  
the \OI\ triplet with O abundances resulting from the forbidden oxygen line
at 6300\,\AA .

As discussed in detail by Allende-Prieto et al. (\cite{prieto01}),
the absorption feature at 6300.3\,\AA\ is a blend of the \oI\ line and a \NiI\ line, 
which makes it slightly  asymmetric. From the best fit of the blend with profiles
calculated from a 3D model atmosphere of the Sun, they derived a solar oxygen
abundance of $A \rm{(O)} _{\odot} \, = \, 8.69$. Later, Johansson et al.
(\cite{johansson03}) measured an experimental oscillator strength of the
\NiI\ line, log\,$gf \, = \, -2.11 \, \pm 0.05$. When using this value and
a solar nickel abundance of $A \rm{(Ni)}_{\odot} = 6.17$ (determined from 17 weak
\NiI\ lines), Scott et al. (\cite{scott09}) obtained  
$A \rm{(O)}_{\odot} \, = \, 8.69$ from a 3D analysis. Repeating the analysis
of the \NiI\ lines with a 1D MARCS model of the Sun,
we obtain a Ni abundance of 6.15 and
predict an equivalent width of 1.7\,m\AA\ for the \NiI\ line in
the $\lambda 6300$ blend. After subtracting this value from the measured
EW of the \oI\ - \NiI\ blend (5.4\,m\AA ) in the solar flux spectrum, 
a solar oxygen abundance of 8.68 is obtained, 
which agrees well with the 3D value.

For the stars we have first calculated the EW of the blending \NiI\ line
using Ni abundances
corresponding to the [Ni/Fe] values of Adibekyan et al. (\cite{adibekyan12}) for 
the HARPS-FEROS sample and those of Nissen \& Schuster (\cite{nissen10}) for
the UVES-FIES sample. The measured EW of the \oI\ - \NiI\ blend is then corrected
for the contribution from the \NiI\ line before the oxygen abundance is determined.

A comparison of non-LTE oxygen abundances derived from the \OI -triplet and O
abundances from the \oI\ line is shown in Fig. \ref{fig:OHcompare.vsTeff-feh}. 
The average of $\Delta \oh \, = \, \oh _{7774} - \oh _{6300}$ 
is 0.010\,dex with an rms scatter of
$\pm 0.07$\,dex, which corresponds well to the errors
estimated in Tables \ref{table:error1} and \ref{table:error2}. As seen,
there is, however, a systematic deviation for the coolest and most metal-rich
stars on the order of $\Delta \oh \sim +0.10$, i.e., the triplet provides higher
oxygen abundances than the $\oI \, \lambda 6300$ line. A similar deviation was found
by Teske et al. (\cite{teske13})
for the planet hosting star \object{55\,Cnc} having \teff = 5350\,K,
\logg = 4.44, and \feh = +0.34. They obtained $\oh _{7774} \simeq$ 
0.22 to 0.27 (depending on which non-LTE corrections were adopted) and
$\oh _{6300} \simeq 0.08$.

\begin{figure}
\resizebox{\hsize}{!}{\includegraphics{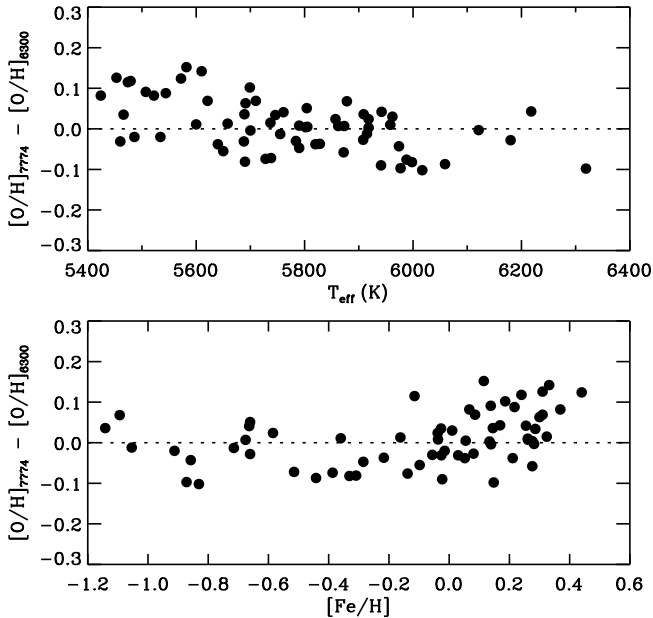}}
\caption{Differences of non-LTE oxygen abundances derived from the
$\OI \, \lambda 7774$ triplet
and oxygen abundances obtained from the $\oI \, \lambda 6300$ line versus \teff\ and \feh .}
\label{fig:OHcompare.vsTeff-feh}
\end{figure}

There are several possible explanations of the discrepancy between \oh\
derived from the \OI\ triplet and the $\oI \, \lambda 6300$ line.
Given that \nife\ increases slightly and \ofe\ decreases with increasing \feh ,
the \NiI\ line makes up a large fraction (up to 60\%)
of the \oI\ - \NiI\ blend in spectra of metal-rich stars, whereas 
the corresponding fraction in the solar spectrum is $\sim$\,30\,\%. These estimates
are based on log\,$gf = -2.11$ for the \NiI\ line as determined
by Johansson  et al. (\cite{johansson03}).
If log\,$gf$ is decreased by for example 0.15\,dex, the importance of the \NiI\ line
decreases, and the solar oxygen abundance derived from the $\oI \, \lambda 6300$ line
increases from $A$(O) = 8.68 to 8.74.  For the cool, metal-rich stars in our sample
the effect is even larger resulting in an increase in $\oh _{6300}$ 
of 0.05\,dex to 0.10\,dex, which removes most of the discrepancy seen
in Fig. \ref{fig:OHcompare.vsTeff-feh}. A decrease in log\,$gf$ of
0.15\,dex is, however, three times larger than the error given by
Johansson et al. (\cite{johansson03}). A lower $gf$-value
would, on the other hand, decrease the puzzling
difference between oxygen abundances derived from the \oI\ lines at 
6300 and 6363\,\AA\ in spectra of dwarf stars (Caffau et al. \cite{caffau13}). 
Hence, we think that this possibility should be kept open.

One may ask if problems with the non-LTE corrections for the \OI\ triplet
could lead to too high \oh\ values compared to those derived from the
$\oI \, \lambda 6300$ line. The corrections of Fabbian et al. (\cite{fabbian09})
have a strong dependence of \teff\ leading to {\em positive} corrections
of \oh\ for dwarf stars cooler than the Sun. If no 
corrections are applied, i.e., LTE is assumed, the discrepancy 
shown in Fig. \ref{fig:OHcompare.vsTeff-feh} for the cool, metal-rich
stars in fact disappears, but instead there is a clear discrepancy for 
stars with $\teff > 6000$\,K, and the scatter of $\oh _{7774} - \oh _{6300}$ 
increases from 0.07\,dex to 0.09\,dex. So, this is not a viable solution.
There may, however, be other ways to solve the problem. 
Schuler et al. (\cite{schuler04}, \cite{schuler06}) found that oxygen
abundances derived from the \OI\ triplet show a strong rise with decreasing
effective temperature for dwarf stars having $\teff < 5450$\,K in the
Pleiades and Hyades open clusters. At $\teff = 5000$\,K there is an overabundance
of $\sim \! 0.25$\,dex relative to the oxygen abundance in the range 
5450\,K $< \teff < 6100$\,K,
and at $\teff = 4500$\,K the overabundance has increased to $\sim 0.75$\,dex.  
Schuler et al. (\cite{schuler06}) suggest that this may be related to the
presence of hot and cool spots on the stellar surface with the hot spots making
a relative large contribution to the EWs of the \OI\ triplet in the cool Pleiades
and Hyades stars. This effect may play a role for the coolest
stars in our sample by contributing to the overabundance of
$\oh _{7774}$ relative to $\oh _{6300}$, although our stars are on average 
older and probably less active than the Pleiades and Hyades stars.

A difference in oxygen abundances derived from 3D model atmospheres and 1D models
is another potential problem. As discussed by Nissen et al.
(\cite{nissen02}), 3D corrections are more important for 
the $\oI \, \lambda 6300$ line than 
the \OI\ triplet, because the \oI\ line is formed in the upper layers
of the atmosphere, where 3D-effects have the largest impact on the temperature
structure. The dependence of 3D corrections on \teff\ is however small, only 0.03\,dex
for a \teff -increase of 400\,K (Nissen et al. \cite{nissen02}, Table 6)
and goes in the wrong direction to diminish
the discrepancy between $\oh _{7774}$ and $\oh _{6300}$ at low \teff .

We conclude from this discussion that there is no obvious explanation of the 
$\oh _{7774}$ - $\oh _{6300}$ discrepancy for the cool, metal-rich stars in our sample,
and that the problem could be with both $\oh _{7774}$ and $\oh _{6300}$.
In the following section we shall, therefore, discuss results based on
both sets of oxygen abundances.

\section{Results and discussion}
\label{sect:discussion}
In this section, we show how \cfe , \ofe , and \co\ change
as a function of increasing metallicity, \feh , for the four
populations of stars identified in the solar neighborhood,
and discuss how these trends may be explained in terms of Galactic chemical
evolution and differences in origin of the various populations.
We also investigate, if there are any systematic differences in the abundance ratios
between stars hosting planets and stars for which no planets have
been detected, and if there is evidence of a
cosmic scatter in \cfe\ and \ofe\ for solar analog stars
corresponding to the variations in
the volatile-to-refractory element ratio claimed in some recent works
(e.g. Mel\'{e}ndez et al. \cite{melendez09}, Ram\'{\i}rez et al.
\cite{ramirez14}).

\subsection{Populations}
\label{sect:populations}
As seen from Tables \ref{table:res1} and \ref{table:res2}, stars
have been classified into four different populations.
For the HARPS-FEROS sample, the \alphafe - \feh\ diagram of
Adibekyan et al. (\cite{adibekyan13}) is used to divide stars into
thin- and  thick-disk populations depending on whether they lie
below or above the dividing line shown in their Fig. 1. At 
$\feh < -0.2$, there is a gab in \alphafe\ between the two populations,
so a chemical separation
is more clear than a kinematical separation, because
thin- and thick-disk stars have considerable overlap
in the Galactic velocity components. For the more metal-rich stars with
$\feh > -0.2$, there is no gab between the 
thin- and thick-disk sequences in the \alphafe - \feh\ diagram, so
the population classification is less clear. Haywood et al.
(\cite{haywood13}) have, however, shown that the metal-rich, alpha-enhanced stars
are older than the metal-rich thin-disk stars and that they form a smooth
extension of the age-metallicity relation for the metal-poor thick-disk  
stars (see Fig. 10 in Haywood et al. \cite{haywood13}).

Stars in the UVES-FIES sample were classified by Nissen \& Schuster
(\cite{nissen10}). If the total space  velocity with respect to
the LSR is larger than 180\,\kmprs , a star is
considered to be a halo star. As shown in Fig. 1 of  Nissen \& Schuster
(\cite{nissen10}), these stars have a bimodal distribution 
in the \alphafe -\feh\ diagram allowing a classification into 
high- and low-alpha halo stars. The two halo populations are  also well
separated in other abundance ratios such as Na/Fe and Ni/Fe with
the low-alpha stars having abundances ratios similar to those of
dwarf galaxies (Tolstoy et al. \cite{tolstoy09}). 
Furthermore, the two populations have different
kinematical properties and ages, i.e., the low-alpha stars have
larger space velocities and more negative (retrograde) $V$ velocity
components than the high-alpha stars, and they are on average  younger
by 2- 3\,Gyr (Schuster et al. \cite{schuster12}). Altogether, this suggests that
the low-alpha stars have been accreted from dwarf galaxies, whereas
the high-alpha stars may have formed in situ  during a dissipative collapse
of proto-Galactic gas clouds.

The UVES-FIES sample includes 16 stars, which were classified by 
Nissen \& Schuster (\cite{nissen10}) as belonging to the thick disk
based on their kinematics. All of these stars fall above the dividing line 
in the \alphafe - \feh\ diagram of Adibekyan et al. (\cite{adibekyan13}),
so they would also be classified as thick-disk according to their
chemical properties.

\subsection{The \cfe\ and \ofe\ trends}
\label{sect:trends}
The trends of \cfe\ and \ofe\ as a function of \feh\ are shown
in Figs. \ref{fig:CFe} and \ref{fig:OFe}, respectively, with 
different symbols for the four populations discussed.
In  Fig. \ref{fig:OFe} we have applied the average oxygen
abundance derived from the \OI\ triplet and the $\oI \, \lambda 6300$
line if both lines were measured. Individual error bars are
given; they are relatively large in \cfe\ for the low-alpha
halo stars due to the weakness of the \CI\ lines.

\begin{figure}
\resizebox{\hsize}{!}{\includegraphics{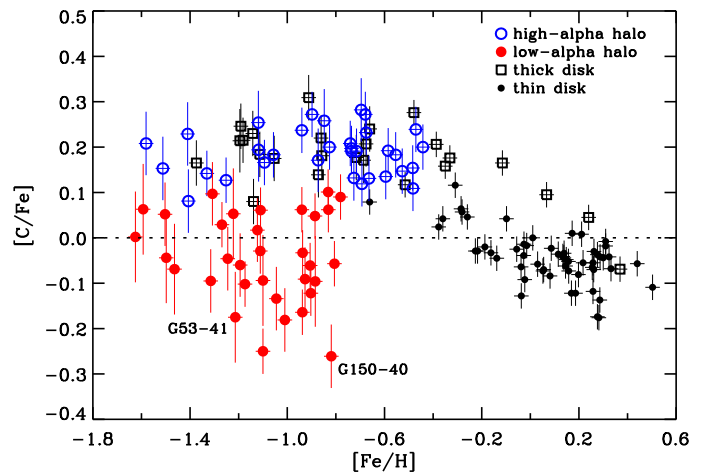}}
\caption{\cfe\ versus \feh\ for stars analyzed in this paper.
The two Na-rich halo stars are marked.}
\label{fig:CFe}
\end{figure}

As seen from  Figs. \ref{fig:CFe} and \ref{fig:OFe}, there
are systematic differences between
thin- and thick-disk stars. The group of thin-disk stars
around $\feh \! \sim \! -0.3$ falls below the thick-disk sequence
in both \cfe\ and \ofe , and the thin-disk star
\object{HD\,203608} at $\feh = -0.66$ also falls below. Furthermore,
we note that the two metal-rich, alpha-enhanced stars
with \feh\ between $-0.15$ and +0.15 clearly fall above the
thin-disk sequence, and lie on an extension of the
more metal-poor thick-disk sequence.
The third and the fourth metal-rich thick-disk stars with
\feh\ between +0.2 and +0.4  do not deviate significantly
from the thin-disk stars
suggesting that the two populations merge in \cfe\ and \ofe\
at the highest metallicities.

The  high-alpha halo stars with $\feh < -0.8$ are distributed around a 
plateau of $\cfe \simeq 0.2$ in Fig. \ref{fig:CFe} and 
$\ofe \simeq 0.6$ in Fig. \ref{fig:OFe}, whereas the low-alpha
stars are shifted downwards with a decreasing trend as  a function of
increasing metallicity. In the case of \ofe , this confirms the
result of Ram\'{\i}rez et al. (\cite{ramirez12}), although their
\ofe\ values tend to be slightly lower. Furthermore, the trends of
\cfe\ and \ofe\ resemble that of \mgfe\ shown in  Nissen \& Schuster
(\cite{nissen10}) except for a constant offset; the high-alpha halo stars
have a plateau at $\mgfe \simeq 0.35$. 

\begin{figure}
\resizebox{\hsize}{!}{\includegraphics{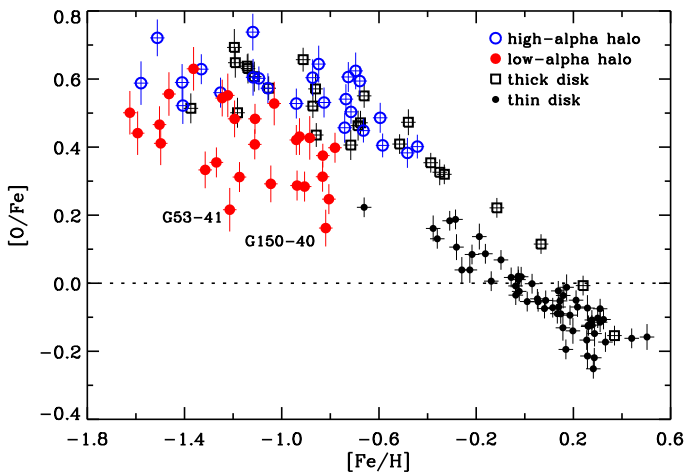}}
\caption{\ofe\ versus \feh . For stars with both $\oh_{6300}$ and
$\oh_{7774}$ available, the average oxygen abundance has been applied.}
\label{fig:OFe}
\end{figure}

Figure \ref{fig:CFe-OFe} shows that 
\cfe\ and \ofe\ are well correlated and that the amplitude
of the variations is about the same for the two
abundance ratios, i.e., the data are well fitted with a line
having a slope of one. Interestingly, the two stars with the
lowest \cfe\ and \ofe\ values, \object{G\,53-41} and 
\object{G\,150-40}, are Na-rich (Nissen \& Schuster \cite{nissen10}).
As discussed by Ram\'{\i}rez et al. (\cite{ramirez12}), they share the
Na-O abundance anomaly of second generation stars born in
globular clusters (Carretta et al. \cite{carretta09}). 

The different trends of high- and low-alpha stars
in Figs. \ref{fig:CFe} and \ref{fig:OFe}
may be explained if the two populations were formed in systems with
different star formation rates. According to this scenario,
the high-alpha stars formed in regions with such a high star formation 
rate that mainly massive stars exploding as 
Type II SNe contributed to the chemical enrichment up to
$\feh \simeq -0.6$ at which  metallicity Type Ia SNe started
to contribute Fe causing \ofe\ to decrease.
The low-alpha stars, on the other hand, originate in
regions with a slow chemical evolution so that 
Type Ia SNe started to contribute iron at
$\feh \simeq -1.6$, or at an even lower metallicity.
Bursts of star formation, as those shown in dwarf galaxies,
could explain the low \cfe\ and \ofe\ present in low-alpha stars.
These values are even lower, if Fe occurs between consecutive
bursts (Carigi et al. \cite{carigi02}).
The two Na-rich stars, \object{G\,53-41} and \object{G\,150-40}, are 
exceptions as they have probably been formed in
globular clusters as mentioned above.

As an alternative explanation of low-alpha 
halo stars, Kobayashi et al. (\cite{kobayashi14})
suggest that they were formed
in regions where the nucleosynthesis contribution
of massive $M > 25 M_{\odot}$ core collapse supernovae 
is missing due to stochastic variations of the IMF.
These stochastic effects are more important in
dwarf galaxies, due to the low content of available gas to form stars 
(Carigi \& Hernandez \cite{carigi08}).
This explanation is based on  the yield calculations of
Kobayashi et al. (\cite{kobayashi06}) showing that the O/Fe
and $\alpha$/Fe ratio is significantly lower for 
13 - 25 $M_{\odot}$ supernovae than in the case of
$M > 25 M_{\odot}$ supernovae. The hypothesis
also explains why Mn/Fe is not enhanced in the low-alpha
stars (Nissen \& Schuster \cite{nissen11}),
which one might have expected if Type Ia SNe contribution of
iron-peak elements is the explanation of the low-alpha
stars. It is, however, difficult to see,  how this
hypothesis can explain why the difference in \cfe ,
\ofe , and \alphafe\ between high- and low-alpha stars
increases as a function of increasing metallicity,
which fact has a natural explanation with a delayed
contribution of Fe from Type Ia SNe.

\begin{figure}
\resizebox{\hsize}{!}{\includegraphics{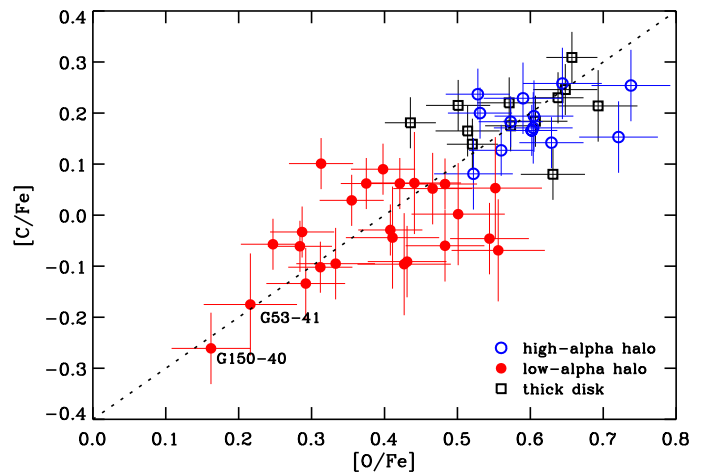}}
\caption{\cfe\ versus \ofe\ for stars with $\feh < -0.8$. 
For stars with both $\oh_{6300}$ and
$\oh_{7774}$ available, the average oxygen abundance has been applied.}
\label{fig:CFe-OFe}
\end{figure}

As seen from  Figs. \ref{fig:CFe} and \ref{fig:OFe},
thick-disk stars show the same trends of \cfe\ and \ofe\ as 
high-alpha halo stars. 
It is not possible to distinguish 
between these two populations from a chemical point of view.
The thick-disk stars differ in kinematics from the high-alpha 
halo stars by having total space velocities,
$\Vtotal < 120$\,\kmprs\ with respect to the LSR,
whereas the high-alpha halo stars were selected to have 
$\Vtotal > 180$\,\kmprs . It could be that both populations have 
been formed during a dissipative
collaps of proto-Galactic gas clouds with each new generation
of stars formed in increasingly flatter and more rotationally
supported spheroids. 

\subsection{The \co\ - \oh\ diagram}
\label{sect:CO-diagram}
The \co \,-\,\oh\ diagram shown in 
Fig. \ref{fig:CO} is particular useful when
discussing the origin and Galactic evolution
of carbon. Given that oxygen is exclusively produced in massive stars
on a relatively short timescale, $\sim \! 10^7$\,yr,
the change in \co\ as a function of \oh\ depends
on the yields and timescales of carbon production
in various types of stars (Chiappini et al. \cite{chiappini03};
Akerman et al. \cite{akerman04};
Carigi et al. \cite{carigi05}; Cescutti et al. \cite{cescutti09};
Carigi \& Peimbert \cite{carigi11}).

As seen from Fig. \ref{fig:CO}, there is no systematic
shift in \co\ between the high- and low-alpha halo populations;
the stars distribute around a 
plateau of $\co \simeq -0.45$ when $\oh < -0.4$.
This result came as a surprise. Assuming that the low \cfe\
and \ofe\ in low-alpha stars are due to Fe
produced by Type Ia SNe, we had expected that low- and
intermediate-mass AGB stars also had enough time to contribute
carbon and raise \co\ in the low-alpha stars
to higher values than in high-alpha stars.
The explanation may be that intermediate-mass (4 - 8 $M_{\odot}$)
AGB stars contribute very little to carbon
(Kobayashi et al.  \cite{kobayashi11}), and
that the evolution timescale of low-mass (1 - 3 $M_{\odot}$) AGB stars,
which do have a high
carbon yield according to Kobayashi et al., is longer than the
timescale for enriching the low-alpha stars with Fe from Type Ia SNe.
Hence, it seems that carbon in high- and low-alpha
halo stars as well as thick-disk stars with $\oh < -0.4$
was made mainly in high-mass stars ($M > 8 \,  M_{\rm Sun}$)
with a C/O yield ratio corresponding to $\co \simeq -0.45$.

\begin{figure}
\resizebox{\hsize}{!}{\includegraphics{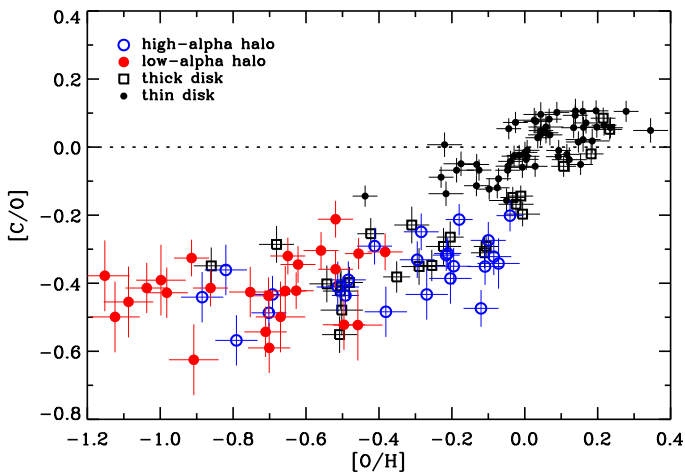}}
\caption{\co\ versus \oh . For stars with both $\oh_{6300}$ and
$\oh_{7774}$ available, the average oxygen abundance has been applied.}
\label{fig:CO}
\end{figure}

For the high-alpha halo and thick-disk stars, \co\ begins to rise
at $\oh \sim -0.4$ and reaches $\co \sim  0.0$ at $\oh \sim +0.2$.
According to Cescutti et al. (\cite{cescutti09}),
this increase can be explained as due to a metallicity dependent 
carbon yield of massive stars, but Akerman et al. (\cite{akerman04})
and Carigi et al. (\cite{carigi05}) ascribe the rise of \co\
to low-mass (1 - 3 $M_{\odot}$) AGB stars of low metallicity.
The thin-disk stars lie on a
different \co\ sequence shifted by $\sim \! 0.2$\,dex to higher \co\ values.
To explain this,  carbon produced in both low-mass
and massive stars has to be included. The 
model by Carigi et al. (\cite{carigi05}) suggests that approximately
half of the carbon in the more metal-rich thin-disk stars has come
from low-mass stars and half from massive stars. A similar conclusion is reached by
Cescutti et al. (\cite{cescutti09}). Still, it remains to be seen
if chemical evolution models can reproduce the observed \co\ trends in detail.

The offset between thin- and thick-disk stars in the \co\ - \oh\ diagram
has previously been found by Bensby \& Feltzing (\cite{bensby06})
in a study where the C and O abundances were derived from the 
forbidden \cI\ $\lambda 8727$ and \oI\ $\lambda 6300$ lines.
Here it is confirmed with C and O abundances based on high-excitation
permitted lines.

\subsection{The C/O ratio in stars with planets}
\label{sect:planets}

As mentioned in the introduction, there is much interest in the C/O
ratio of stars hosting planets, because this ratio may have important
effects on the structure and composition of the planets. High
ratios (C/O $> 0.8$) were found in a significant fraction (10-15\,\%)
of F and G dwarf stars by Delgado Mena et al. (\cite{delgado10})
and Petigura \& Marcy (\cite{petigura11}) based on oxygen abundances
derived from the forbidden \oI\ $\lambda 6300$ line, but this
was not confirmed by Nissen (\cite{nissen13}), who derived 
oxygen abundances from the \OI\ triplet at 7774\,\AA .

\begin{figure}
\resizebox{\hsize}{!}{\includegraphics{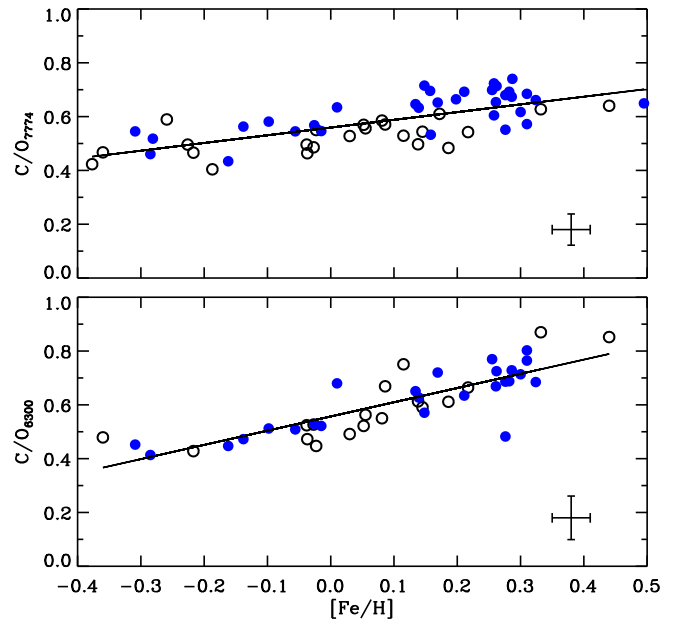}}
\caption{The C/O ratio versus \feh\ for thin-disk stars. Stars proven
to have planets (see Table \ref{table:res1}) are plotted with filled 
(blue) circles. In the upper panel, 
O abundances have been determined from the \OI\ triplet and in the 
lower panel from the \oI\ $\lambda 6300$ line. The lines show 
the relations in Eqs. (2) and (3), respectively}
\label{fig:CO.ratios}
\end{figure}

In this paper we have 
included a number of stars with  no detection of planets.
C/O as a function of \feh\ for thin-disk stars is shown in Fig. \ref{fig:CO.ratios}
with oxygen abundances derived from either the  \OI\ triplet at 7774\,\AA\
(upper panel) or the \oI\ $\lambda 6300$ line (lower panel). From linear
regression, $y \, = \, A \, + \, B\,x$, we obtain
\begin{eqnarray}
{\rm C/O}_{7774} = 0.559 \, (\pm 0.009) + 0.287 \, (\pm 0.039) \,\,\feh 
\end{eqnarray}
with a standard deviation of $\pm 0.061$ and 
\begin{eqnarray}
{\rm C/O}_{6300} = 0.557 \, (\pm 0.012) + 0.529 \, (\pm 0.055) \,\,\feh 
\end{eqnarray}
with a standard deviation of $\pm 0.068$.

In both cases, the zero-point (at \feh\ = 0.0) is 
close to the solar value, (C/O)$_{\odot} = 0.58$, determined  
from HARPS-FEROS spectra of reflected sunlight from Ceres and 
Ganymede. The slope of C/O$_{6300}$ is, however, higher than that
of C/O$_{7774}$, which is connected to the systematic differences
between oxygen abundances derived from the \OI\ triplet and 
the forbidden line discussed in Sect. \ref{sect:COabundances}.
In any case, both fits correspond to C/O values  below 0.8
up to a metallicity of $\feh  \, \simeq +0.4$, and only two stars have  C/O$_{6300}$ 
above 0.8, which may be statistical deviations.
Hence, we confirm the conclusion of Nissen (\cite{nissen13})
that there is no evidence of the existence of stars with C/O
values above 0.8, i.e., the critical limit for the formation of 
carbon planets (Bond et al.  \cite{bond10}). 

The dispersions in C/O around the fitted lines in  Fig. \ref{fig:CO.ratios}
correspond well to the errors of \co\ estimated in Table \ref{table:error1}.
Stars with planets detected\footnote{Based on the April 2014 version of 
the Exoplanet Orbit Database at {\tt http://exoplanets.org} described by
Wright et al. (\cite{wright11})} tend, however, to lie 
above the fitted lie in C/O$_{7774}$; if only stars with planets are
included in the regression, the zero-point of the fit 
becomes $A_{7774} = 0.586 \pm0.013$ compared
to $A_{7774} = 0.525 \pm0.013$ for stars without
planets. This difference is statistically significant
at a 3-sigma level. On the other hand, a smaller difference is found
in the case of C/O$_{6300}$. Here, 
the zero-points of the fits are $A_{6300} = 0.562 \pm0.016$ for stars
with planets and $A_{6300} = 0.552 \pm0.017$ for stars without planets.

\subsection{Scatter of \cfe\ and \ofe\ among solar analog stars}
\label{sect:scatter}
In a recent high-precision abundance work, Ram\'{\i}rez et al. (\cite{ramirez14})
found the slope of abundance ratios [X/Fe] versus condensation temperature, 
$T_{\rm C}$, of element X to vary with 
an amplitude of $\sim \! 10^{-4}$\,dex\,K$^{-1}$. These variations were found 
for late F-type dwarfs, metal-rich solar analogs, and solar-twin stars.
While the 50\% condensation temperatures of C and O are 40 and 180\,K, 
respectively, it is 1330\,K for Fe in a solar-system composition gas
(Lodders \cite{lodders03}). Hence, the results of Ram\'{\i}rez et al.
imply that \cfe\ and \ofe\ should vary
with an amplitude of approximately 0.1\,dex, which magnitude may be
detectable as a cosmic scatter in the \cfe\ and \ofe\ trends for
our sample of F and G dwarfs stars. Furthermore, the study of 
Mel\'{e}ndez et al. (\cite{melendez.etal09}) showed the 
Sun to have a high volatile-to-refractory ratio (corresponding to 
$\cfe \, \sim \ofe \, \sim \,0.05$) relative to the ratio 
in 11 solar twins, which they suggest
may be due to depletion of refractory elements when 
terrestrial planets formed.

In order to test these intriguing results, we have selected
a sample of 40 solar ``analog" stars, defined as thin-disk
stars with effective temperatures within $\pm 200$\,K from the solar 
\teff\ and metallicities in the range
$-0.4 < \feh < +0.4$. The surface gravity of this sample ranges
from \logg = 3.90 to 4.62, but the large majority ($N = 32$) of the stars
have $\logg > 4.20$. The means of the atmospheric parameters are  
$<\teff > = 5760$\,K and $<\logg > = 4.35$, i .e. close to the
solar parameters.

Figure \ref{fig:CFe.analogs} 
shows a plot of \cfe\ versus \feh\ for these solar analogs.
From linear regression we obtain
\begin{eqnarray}
\cfe = -0.029 \, (\pm 0.008) - 0.151 \, (\pm 0.035) \,\,\feh
\end{eqnarray}
with a standard deviation $\sigma \cfe = 0.047$.

The upper panel of Fig. \ref{fig:OFe.analogs} shows
the relation between $\ofe_{7774}$ and \feh\ for the
sample of 40 solar analog stars and in the lower panel 
$\ofe_{6300}$ versus \feh\ is shown for the subsample of 32
solar analogs, which have O abundances determined from the \oI\
$\lambda 6300$ line. The linear regressions are
\begin{eqnarray}
\ofe_{7774} = -0.003 \, (\pm 0.006) - 0.399 \, (\pm 0.029) \,\,\feh
\end{eqnarray}
with a standard deviation $\sigma \ofe_{7774} = 0.039$, and
\begin{eqnarray}
\ofe_{6300} = -0.006 \, (\pm 0.009) - 0.533 \, (\pm 0.045) \,\,\feh
\end{eqnarray}
with $\sigma \ofe_{6300} = 0.047$.

\begin{figure}
\resizebox{\hsize}{!}{\includegraphics{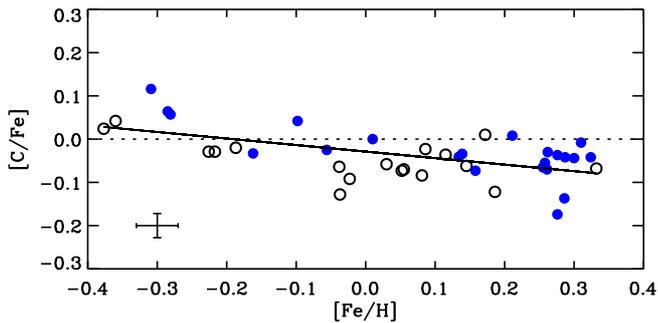}}
\caption{\cfe\ versus \feh\ for solar analog stars.
Stars proven to have planets are plotted with filled (blue) circles. 
The line shows the relation in  Eq. (4).}
\label{fig:CFe.analogs}
\end{figure}

\begin{figure}
\resizebox{\hsize}{!}{\includegraphics{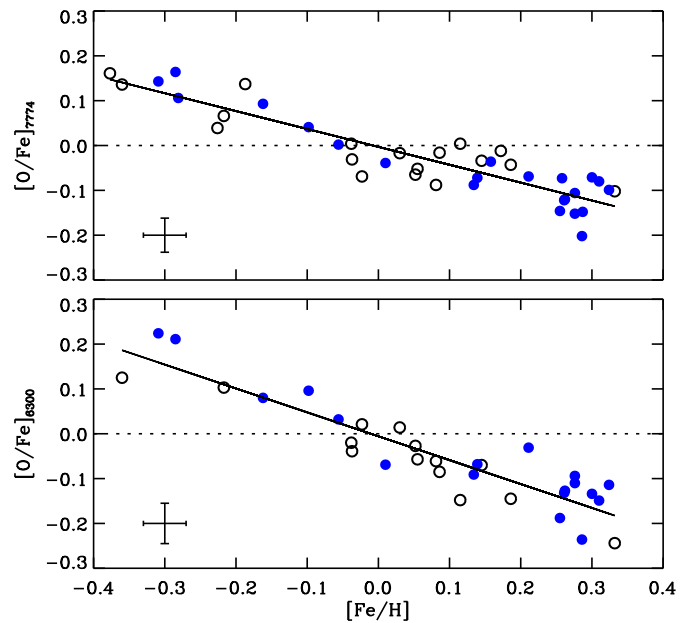}}
\caption{\ofe\ versus \feh\ for solar analog stars. Stars proven
to have planets are plotted with filled (blue) circles. In the upper panel, 
O abundances have been determined from the \OI\ triplet and in the
lower panel from the \oI\ $\lambda 6300$ line. The lines correspond
to Eqs. (5) and (6), respectively}
\label{fig:OFe.analogs}
\end{figure}

For $\ofe_{7774}$ and $\ofe_{6300}$, the standard deviations 
agree well with the errors estimated in Table \ref{table:error1},
but in the case of \cfe\ the standard deviation of the regression
($\pm 0.047$\,dex) is higher that the estimated error ($\pm 0.028$\,dex).
We note, in this connection, that an extension of the 
linear regression to include
terms in \teff\ and \logg\ does not decrease the standard deviation
significantly (from $\pm 0.047$\,dex to $\pm 0.044$\,dex).

As seen from Figs. \ref{fig:CFe.analogs} and \ref{fig:OFe.analogs},
stars with metallicities close to that of the Sun tend to fall below
the fits suggesting that {\em linear} regression is not providing the
right mean values of \cfe\ and \ofe\ for stars with $\feh \simeq 0.0$.
For the group of 11 stars with $-0.1 < \feh < +0.1$\footnote{The average values
of the atmospheric parameters of these stars,
$<\teff>\, = 5808$\,K, $<\logg>\, = 4.43$, and $<\feh>\, = 0.005$, are
close to the solar values.}, the average values
are $<\!\cfe \!>\, = -0.052$, $<\!\ofe _{7774}\!>\, = -0.029$, 
and $<\!\ofe _{6300}\!>\, = -0.016$.
The small offsets of \ofe\ may be explained as due to accidental errors
in the measured equivalent widths of the \OI\ triplet and 
\oI\ $\lambda 6300$ lines in the solar spectrum, but to explain the offset
in \cfe\ in this way, the solar EWs of the \CI\ lines at 5052\,\AA\ and 
5380\,\AA\ would have to be overestimated by more than 2\,m\AA . This is
unlikely to be the case, because the HARPS spectrum of the Sun was obtained
from reflected sunlight of Ceres and Ganymede in the same way
as the stellar spectra. The spectra of these minor planets have S/N ratios
of 350 and 250, respectively, and the EWs in the two spectra agree within
0.7\,m\AA\ for the $\lambda$\,5052 line and within 0.2\,m\AA\
for the $\lambda$\,5380 line.
We also note that the stellar \teff -scale at $\feh \simeq 0$
has to be wrong by $\sim$\,150\,K
to explain the offset of \cfe , which is much larger than the 
error of $\pm 15$\,K (estimated from solar twin stars)
of the Cassagrande et al. (\cite{casagrande10}) 
IRFM calibration.
Hence, it seems that the difference of \cfe\ between the Sun 
and stars with solar metallicity is real. Furthermore, we note 
that stars without detected planets in Fig. \ref{fig:CFe.analogs}
tend to fall below stars proven to have planets.
Fitting the two samples separately, we get
\begin{eqnarray}
\cfe = -0.003 \, (\pm 0.011) - 0.215 \, (\pm 0.047) \,\,\feh
\end{eqnarray}
for 22 stars with planets, and
\begin{eqnarray}
\cfe = -0.051 \, (\pm 0.011) - 0.132 \, (\pm 0.059) \,\,\feh
\end{eqnarray}
for 18 stars without planets. The difference in zero-points is
significant at the 3-sigma level, and it is intriguing that
stars with planets have the same \cfe\ at $\feh \simeq 0.0$ as
the Sun.

It is tempting to think that
the scatter in \cfe\ and the possible systematic difference
in \cfe\ between stars with and without planets 
are related  to the
variations in the volatile-to-refractory ratio found in
the works of Mel\'{e}ndez et al. (\cite{melendez09}) and
Ram\'{i}rez et al. (\cite{ramirez14}). However, 
similar ``cosmic" variations are not seen for \ofe ;
there is no significant difference in the trends of \ofe\ for 
stars with and without planets, and the scatter around
the fitted lines is not larger than expected from the estimated error of \ofe .
This seems to rule out
that the variations in \cfe\ (if real) can be explained as due to
variations in the volatile-to-refractory ratio.

\section{Conclusions}
\label{sect:conclusions}
In this paper, we have determined precise C and O abundances
for F and G main-sequence stars in the solar neighborhood
belonging to four different
populations defined from their sequences in the 
\alphafe\ -- \feh\ diagram: thin- and thick-disk stars, 
high- and low-alpha halo stars. Based on the trends of
\cfe , \ofe , and \co , we make the following conclusions.

i) The high- and low-alpha halo populations are clearly separated
in \cfe\ and \ofe\ for the metallicity range $-1.6 < \feh < -0.8$
(see Figs. \ref{fig:CFe} and \ref{fig:OFe}). 
The  high-alpha stars have approximately constant values of
$\cfe \simeq 0.2$ and $\ofe \simeq 0.6$, whereas the low-alpha
stars show decreasing trends in \cfe\ and \ofe\ as  a function of
increasing metallicity and present a higher dispersion in
these abundance ratios than the high-alpha stars. These chemical
properties and the kinematics of the stars
can be explained in a scenario, where
the high-alpha stars formed in situ in a dissipative collapse
of proto-Galactic gas clouds, where mainly massive stars exploding as
Type II SNe contributed to the chemical enrichment.
The low-alpha stars, on the other hand, may have been accreted from dwarf galaxies
with a relatively slow and bursting star formation history, 
so that Type Ia SNe started to contribute iron at $\feh \simeq -1.6$.

ii) For metallicities below $\feh \simeq -0.4$,  stars with
thick-disk kinematics follow the same trends in
\cfe\ and \ofe\ as the high-alpha halo stars. In both populations,
\ofe\ begins to decline at $\feh \simeq -0.6$
because Type Ia SNe start to contribute iron at 
this metallicity. This suggest a similar origin of thick-disk and
high-alpha halo stars, possibly during the dissipative collaps of
proto-Galactic gas clouds.

iii) Although only a few metal-rich thick-disk stars are
included in this study, we find clear evidence that
thin- and thick disk stars (classified from \alphafe )
are separated in \cfe\ and \ofe\ 
in the metallicity range $-0.6 < \feh < +0.1$. This supports
recent evidence of the existence of alpha-enhanced, metal-rich
stars (Adibekyan et al. \cite{adibekyan12}; Bensby et al. \cite{bensby14})
that are older than the thin-disk stars (Haywood et al. \cite{haywood13}). 

iv) The \co\ - \oh\ diagram (Fig. \ref{fig:CO}) shows no offset in \co\
between the high- and low-alpha halo stars, which suggests that mainly
high-mass ($M > 8 M_{\odot}$) stars have contributed C and O in these populations. 
The rise in \co\ at $\oh > -0.4$ for the high-alpha and thick-disk
populations may be due to a metallicity dependent carbon yield of  
such high-mass stars (Cescutti et al. \cite{cescutti09}) and/or
the C contribution from low-mass stars (Carigi et al. \cite{carigi05}). 
The thin-disk stars 
show an offset in \co\ relative to the thick-disk stars
due to additional  C contribution from low-mass AGB star
and massive stars of high metallicity (Carigi et al. \cite{carigi05};
Carigi \& Peimbert \cite{carigi11}), which is made possible due to the 
younger age of the thin-disk population.

v) The C/O ratio of thin-disk stars (Fig. \ref{fig:CO.ratios}) shows a  
tight, increasing trend as a function of \feh , but even at the
highest metallicities, $\feh \simeq +0.4$, the ratio does not exceed 0.8,
i.e., the critical value in a proto-planetary disk
for the formation of carbon planets according
to Bond et al. (\cite{bond10}). This confirms recent results obtained by
Nissen (\cite{nissen13}) and Teske et al. (\cite{teske14}).
There is some evidence of higher C/O ratios in stars hosting planets
than in stars without detected planets, but this needs to be
confirmed.

vi) In a sample of solar analog stars, there is 
evidence of a small ($0.05\,\pm 0.016$\,dex) systematic difference in \cfe\
between stars with and without detected planets. An attempt
to explain this in terms of variations in the volatile-to-refractory
element ratio (Mel\'{e}endez et al. \cite{melendez09}:
Ram\'{\i}rez et al. \cite{ramirez14}) fails, because a similar
difference in \ofe\ between stars with and without planets
is not found.

\begin{acknowledgements}
This project is supported by the National Natural Science
Foundation of China through Grant No. 11390371.
P.E.N. acknowledges a visiting professorship
at the National Astronomical Observatories in Beijing granted
by the Chinese Academy of Sciences (Contract no. 6-1309001).
L.C. thanks for the financial supports provided by CONACyT of Mexico
(grant 129753) and by MINECO of Spain (AYA2010-16717 and AYA2011-22614).
Funding for the Stellar Astrophysics Centre is provided by the
Danish National Research Foundation (Grant agreement no.: DNRF106).
The research is supported by the ASTERISK project
(ASTERoseismic Investigations with SONG and Kepler)
funded by the European Research Council (Grant agreement no.: 267864).
This research made use of the SIMBAD database operated
at CDS, Strasbourg, France, the Exoplanet Orbit Database and the
Exoplanet Data Explorer at {\tt http://exoplanets.org},
and of data products from the Two Micron All
Sky Survey, which is a joint project of the University of Massachusetts and
the Infrared Processing and Analysis Center/California Institute of
Technology, funded by NASA and the National Science Foundation. 
\end{acknowledgements}

\Online


\begin{thebibliography}{}

\bibitem[2011]{adibekyan11}
Adibekyan, V. Zh., Santos, N. C., Sousa, S. G., \& Israelian, G.  2011, A\&A, 535, L11

\bibitem[2012]{adibekyan12}
Adibekyan, V. Zh., Sousa, S. G. , Santos, N. C., et al. 2012, A\&A, 545, A32

\bibitem[2013]{adibekyan13}
Adibekyan, V. Zh., Figueira, P., Santos, N. C., et al. 2013, A\&A, 554, A44 

\bibitem[2004]{akerman04}
Akerman, C. J., Carigi, L., Nissen, P. E., Pettini, M., \& Asplund, M. 
2004, A\&A, 414, 931

\bibitem[2001]{prieto01}
Allende Prieto, C., Lambert, D. L., \& Asplund, M. 2001, ApJ, 556, L63


\bibitem[2005]{asplund05}
Asplund, M. 2005, ARA\&A, 43, 481


\bibitem[2009]{asplund09}
Asplund, M., Grevesse, N., Sauval, A. J., \& Scott, P. 2009, ARA\&A, 47, 481

\bibitem[2007]{barklem07}
Barklem, P. S. 2007, A\&A, 462, 781

\bibitem[2005]{barklem05}
Barklem, P. S. \& Aspelund-Johansson, J. 2005, A\&A, 435, 373

\bibitem[2000]{barklem00}
Barklem, P. S., Piskunov, N., \& O'Mara, B. J. 2000, A\&AS, 142, 467

\bibitem[2006]{bensby06}
Bensby, T., \& Feltzing, S. 2006, MNRAS, 367, 1181

\bibitem[2004]{bensby04}
Bensby, T., Feltzing, S., \& Lundstr{\"o}m, I. 2004, A\&A, 415, 155

\bibitem[2005]{bensby05}
Bensby, T., Feltzing, S., Lundstr{\"o}m, I., Ilyin, I. 2005, A\&A, 433, 185

\bibitem[2014]{bensby14}
Bensby, T., Feltzing, S., Oey, M. S. 2014, A\&A, 562, A71  

\bibitem[2013]{bond13}
Bond, H. E., Nelan, E. P., VandenBerg, D. A., Schaefer, G. H., \&
Harmer, D. 2013, ApJ, 765, L12

\bibitem[2010]{bond10}
Bond, J. C., O'Brien, D. P., \& Lauretta, D. S. 2010, ApJ, 715, 1050

\bibitem[2010]{caffau10}
Caffau, E., Ludwig, H.-G., Bonifacio, P., et al. 2010,  A\&A, 514, A92

\bibitem[2013]{caffau13}
Caffau, E., Ludwig, H.-G., Malherbe, J.-M., et al. 2013,  A\&A, 554, A126

\bibitem[2008]{caffau08}
Caffau, E., Ludwig, H.-G., Steffen, M., et al. 2008,  A\&A, 488, 1031

\bibitem[2008]{carigi08}
Carigi, L., \& Hernandez, X, 2008, MNRAS, 390, 582

\bibitem[2002]{carigi02}
Carigi, L., Hernandez, X, \& Gilmore, G. 2002, MNRAS, 334, 117

\bibitem[2011]{carigi11}
Carigi, L., \& Peimbert, M. 2011, RMxAA, 47, 139

\bibitem[2005]{carigi05}
Carigi, L., Peimbert, M., Esteban, C., \& Garc\'{\i}a-Rojas, J. 2005, ApJ, 623,  213

\bibitem[2009]{carretta09}
Carretta, E., Bragaglia, A., Gratton, R., \& Lucatello, S. 2009,
A\&A, 505, 139

\bibitem[2010]{casagrande10}
Casagrande, L., Ram\'{\i}rez. I., Mel\'{e}ndez, J., Bessell, M., \& Asplund, M. 2010,
A\&A, 512, A54 

\bibitem[2004]{cayrel04}
Cayrel, R., Depagne, E., Spite, M., et al. 2004, A\&A, 416, 1117

\bibitem[2009]{cescutti09}
Cescutti, G., Matteucci, F., McWilliam, A., \& Chiappini, C. 2009, A\&A, 505, 605

\bibitem[2003]{chiappini03}
Chiappini, C., Matteucci, F., \& Meynet, G. 2003, A\&A, 410, 257

\bibitem[2010]{delgado10}
Delgado Mena, E., Israelian, G., Gonz\'{a}lez Hern\'{a}ndez, J. I., et al. 2010,
ApJ, 725, 2349


\bibitem[1968]{drawin68}
Drawin, H.-W. 1968, Zeitschrift f{\"u}r Physik, 211, 404

\bibitem[2009]{fabbian09}
Fabbian, D., Asplund, M., Barklem, P. S., Carlsson, M., \& Kiselman, D.
2009,  A\&A, 500, 1221

\bibitem[2012]{fortney12}
Fortney, J. J. 2012, ApJ, 747, L27

\bibitem[2003]{fulbright03}
Fulbright, J. P., \& Johnson, J. A. 2003, ApJ, 595, 1154

\bibitem[2006]{garcia06}
Garc\'{\i}a P\'{e}rez, A. E., Asplund, M., Primas, F., Nissen, P. E., \&
Gustafsson, B. 2006, A\&A, 451, 621


\bibitem[2010]{gonzalez10}
Gonz\'{a}lez Hern\'{a}ndez, J. I., Israelian, G., Santos, N. C., et al. 2010
ApJ, 720, 1592

\bibitem[2013]{gonzalez13}
Gonz\'{a}lez Hern\'{a}ndez, J. I., Delgado-Mena, E., Sousa, S. G., et al. 2013
A\&A, 552, A6

\bibitem[2008]{gustafsson08}
Gustafsson, B., Edvardsson, B., Eriksson, K., et al.
2008, A\&A, 486, 951

\bibitem[1999]{gustafsson99}
Gustafsson, B., Karlsson, T., Olsson, E., Edvardsson, B., \& Ryde, N. 1999, A\&A, 342, 426

\bibitem[2013]{haywood13}
Haywood, M., Di Matteo, P., Lehnert, M. D., Katz, D., \& G\'{o}mez, A
2013, A\&A 560, A109

\bibitem[1991]{hibbert91}
Hibbert, A., Bi{\'e}mont, E., Godefroid, M., \& Vaeck, N. 1991, J. Phys. B, 24, 3943

\bibitem[1993]{hibbert93}
Hibbert, A., Bi{\'e}mont, E., Godefroid, M., \& Vaeck, N. 1993, A\&AS, 99, 179      

\bibitem[2003]{johansson03}
Johansson,  S., Litz{\'e}n, U., Lundberg, H., \& Zhang, Z. 2003, ApJ, 584, L107

\bibitem[1999]{kaufer99}
Kaufer, A., Stahl, O., Tubbesing, K., et al. 1999, The Messenger, 95, 8

\bibitem[1993]{kiselman93}
Kiselman, D. 1993, A\&A, 275, 269 

\bibitem[2014]{kobayashi14}
Kobayashi, C., Ishigaki, M. N., Tominaga, N., \& Nomoto, K. 2014
ApJ, 785, L5

\bibitem[2011]{kobayashi11}
Kobayashi, C., Karakas, A. I., \& Umeda, H. 2011,
MNRAS, 414, 3231

\bibitem[2006]{kobayashi06}
Kobayashi, C., Umeda, H., Nomoto, K., Tominaga, N., \& Ohkubo, T.
2006, ApJ, 653, 1145

\bibitem[2005]{kuchner05}
Kuchner, M.J., \& Seager, S. 2005, arXiv:astro-ph/0504214

\bibitem[2012]{lind12}
Lind, K., Bergemann, M., \& Asplund, M. 2012, MNRAS, 427, 50

\bibitem[2012]{liushu12}
Liu, S., Nissen, P. E., Schuster, W. J., et al. 2012, A\&A, 541, A48

\bibitem[2003]{lodders03}
Lodders, K. 2003, ApJ, 591, 1220

\bibitem[2012]{madhusudhan12}
Madhusudhan, N., Lee, K. K. M., \& Mousis, O. 2012, ApJ, 759, L40
 
\bibitem[2011]{mashonkina11}
Mashonkina, L., Gehren, T., Shi, J.-R, Korn, A. J., \& Grupp, F.
2011, A\&A, 528, A87

\bibitem[2003]{mayor03}
Mayor, M., Pepe, F., Queloz, D., et al. 2003, The Messenger , 114, 20

\bibitem[2009]{melendez09}
Mel\'{e}ndez, J. \&  Barbuy, B. 2009, A\&A, 497, 611

\bibitem[2009]{melendez.etal09}
Mel\'{e}ndez, J. Asplund, M., Gustafsson, B., \& Yong, D. 2009, ApJ, 704, L66

\bibitem[2002]{meynet02}
Meynet, G., \& Maeder, A. 2002, A\&A, 390, 561

\bibitem[2013]{nissen13}
Nissen, P. E. 2013, A\&A, 552, A73

\bibitem[2002]{nissen02}
Nissen, P. E., Primas, F., Asplund, M., \& Lambert, D. L. 2002, A\&A, 390, 235

\bibitem[1997]{nissen97}
Nissen, P. E., \& Schuster, W. J. 1997, A\&A, 326, 751

\bibitem[2010]{nissen10}
Nissen, P. E., \& Schuster, W. J. 2010, A\&A, 511, L10

\bibitem[2011]{nissen11}
Nissen, P. E., \& Schuster, W. J. 2011, A\&A, 530, A15

\bibitem[2012]{nissen12}
Nissen, P. E., \& Schuster, W. J. 2012, A\&A, 543, A28

\bibitem[2014]{nissen14}
Nissen, P. E., \& Schuster, W. J. 2014,
in Proc. IAUS298 {\em Setting the scene for GAIA and LAMOST},
eds. S. Feltzing, G. Zhao, N. A. Walton, and P. A. Whitelock,
(Cambridge Univ. Press) p. 65

\bibitem[1983]{olsen83}
Olsen, E. H. 1983, A\&AS, 54, 55

\bibitem[2009]{pereira09}
Pereira, T. M. D., Asplund, M., \& Kiselman, D. 2009,  A\&A, 508, 1403

\bibitem[1997]{perryman97}
Perryman, M. A. C. (ed.) 1997, The HIPPARCOS and TYCHO catalogues, ESA SP-1200

\bibitem[2011]{petigura11}
Petigura, E. A., \& Marcy, G. W. 2011, ApJ, 735, 41

\bibitem[2013]{ramirez13}
Ram\'{\i}rez, I., Allende Prieto, C., \& Lambert, D. L. 2013,
ApJ., 764, 78

\bibitem[2005]{ramirez05}
Ram\'{\i}rez, I., \& Mel\'{e}ndez, J. 2005, ApJ, 626, 465


\bibitem[2012]{ramirez12}
Ram\'{\i}rez, I., Mel\'{e}ndez, J., \& Chanam\'{e}, J. 2012, ApJ, 757, 164

\bibitem[2014]{ramirez14}
Ram\'{\i}rez, I., Mel\'{e}ndez, J., \& Asplund, M. 2014, A\&A, 561, A7

\bibitem[2006]{reddy06}
Reddy, B. E., Lambert, D. L., \& Allende Prieto, C. 2006,
MNRAS, 367, 1329

\bibitem[2004]{schuler04}
Schuler, S. C., King, J. R., Hobbs, L. M., Pinsonneault, M. H. 2004, ApJ, 602, L117

\bibitem[2006]{schuler06}
Schuler, S. C., King, J. R., Terndrup, D. M., et al. 2006, ApJ, 636, 432

\bibitem[2012]{schuster12}
Schuster, W. J., Moreno, E., Nissen, P. E., \& Pichardo, B. 2012
A\&A, 538, A21

\bibitem[2009]{scott09}
Scott, P., Asplund, M., Grevesse, N., \& Sauval, J. 2009, ApJ, 691, L119

\bibitem[2002]{shi02}
Shi, J. R., Zhao, G., \& Chen, Y. Q. 2002, A\&A, 381, 982

\bibitem[2006]{skrutskie06}
Skrutskie, M.~F., Cutri, R.~M., Stiening, R., et al. 2006, AJ, 131, 1163

\bibitem[2008]{sousa08}
Sousa, S. G., Santos, N. C., Mayor, M., et al. 2008, A\&A, 487, 373

\bibitem[2011a]{sousa11a}
Sousa, S. G., Santos, N. C., Israelian, G., et al. 2011a, A\&A, 526, A99

\bibitem[2011b]{sousa11b}
Sousa, S. G., Santos, N. C., Israelian, G., Mayor, M., \& Udry, S. 2011b, A\&A, 533, A141

\bibitem[2005]{takeda05}
Takeda, Y., \& Honda, S. 2005, PASJ, 57, 65

\bibitem[2013]{takeda13}
Takeda, Y, \& Takada-Hidai, M. 2013, PASJ, 65, 65

\bibitem[2013]{teske13}
Teske, J. K., Cunha, K., Schuler, S. C., Griffith, C. A., \& Smith, V. V. 
2013, ApJ, 778, 132

\bibitem[2014]{teske14}
Teske, J. K., Cunha, K., Smith, V. V.,  Schuler, S. C., \& Griffith, C. A.
2014, ApJ, 788, 39 

\bibitem[2009]{tolstoy09}
Tolstoy, E., Hill, V., \& Tosi, M. 2009, \araa , 47, 371

\bibitem[1955]{unsold55}
Uns{\"o}ld, A. 1955, Physik der Sternatmosph{\"a}ren, 2nd ed. (Berlin: Springer Verlag)

\bibitem[1997]{hoek97}
van den Hoek, L. B., \& Groenewegen, M. A. T. 1997, A\&AS, 123, 305

\bibitem[2007]{leeuwen07}
van Leeuwen, F. 2007, Hipparcos, the New Reduction of the Raw Data,
(Astrophys. Space Sci. Library, vol. 350; Dordrecht, Springer)

\bibitem[2011]{wright11}
Wright, J. T., Fakhouri, O., Marcy, G. W., et al. 2011, PASP, 123, 412

\bibitem[2003]{yi03}
Yi, S. K., Kim, Y. -C., \& Demarque, P. 2003, ApJS, 144, 259

\end{thebibliography}
\end{document}